\documentclass[11pt,a4paper,final]{article}
\usepackage{amsmath}%
\numberwithin{equation}{section}
\usepackage{amstext}%
\usepackage{amssymb}%
\usepackage{showkeys}%
\usepackage{epsfig}%
\usepackage{graphicx}%
\usepackage{xcolor}
\usepackage{multicol}
\usepackage{amsthm}
\usepackage{booktabs}
\usepackage{apacite}
\usepackage{multirow}
\usepackage[top=1in, bottom=1.25in, left=0.8in, right=0.8in]{geometry}

\theoremstyle{prop}
\theoremstyle{proof}

\providecommand{\keywords}[1]{
\textbf{Keywords:~~~} Stochastic DDM, ML estimator, Bayesian estimator, Kalman filtering, Regime--switching.
}

\begin{document}
\title{Parameter Estimation Methods of Required Rate of Return}
\author{Battulga Gankhuu\footnote{Department of Applied Mathematics, National University of Mongolia; E-mail: battulga.g@seas.num.edu.mn; Phone Number: 976--99246036}}
\date{}

\maketitle 

\begin{abstract}
In this study, we introduce new estimation methods for the required rate of returns on equity and liabilities of private and public companies using the stochastic dividend discount model (DDM). To estimate the required rate of return on equity, we use the maximum likelihood method, the Bayesian method, and the Kalman filtering. We also provide a method that evaluates the market values of liabilities. We apply the model to a set of firms from the S\&P 500 index using historical dividend and price data over a 32--year period. Overall, the suggested methods can be used to estimate the required rate of returns.
\end{abstract}


\section{Introduction}

Dividend discount models (DDMs), first introduced by \citeA{Williams38}, are a popular tool for stock valuation. If we assume that a firm will not default in the future, then the basic idea of all DDMs is that the market price of a stock equals the sum of the stock's next period price and dividend discounted at a risk--adjusted rate, known as a required rate of return, see, e.g., \citeA{Brealey20}. By their very nature, DDM approaches are best applicable to companies paying regular cash dividends. For a DDM with default risk, we refer to \citeA{Battulga22a}. As the outcome of DDMs depends crucially on dividend forecasts, most research in the last few decades has been around the proper estimations of dividend development. An interesting review of some existing deterministic and stochastic DDMs, which model future dividends can be found in \citeA{dAmico20a}. 

It is an obvious fact that in addition to dividend forecast models, the required rate of return is the main input parameter for DDMs. In addition to its usage in stock valuation, it is an ingredient of the weighted average cost of capital (WACC), and WACC is used to value businesses and projects, see \citeA{Brealey20}. The most common model to estimate the required rate of return is the capital asset pricing model (CAPM). Using the CAPM is common in practice, but it is a one--factor model ($\beta$ only) for which criticism applies, see, e.g., \citeA{Nagorniak85}. Thus, multi--factor models (e.g., \citeA{Fama93}) are therefore often preferred instead. Another multi--factor model, which is used to estimate the required rate of return is \citeauthor{Ross76}'s \citeyear{Ross76} arbitrage pricing theory (APT). However, for the model, since every analyst can develop his APT model, there is no universally accepted APT model specification among practitioners. 

Sudden and dramatic changes in the financial market and economy are caused by events such as wars, market panics, or significant changes in government policies. To model those events, some authors used regime--switching models. The regime--switching model was introduced by seminal works of \citeA{Hamilton89,Hamilton90} (see also a book of \citeA{Hamilton94}) and the model is a hidden Markov model with dependencies, see \citeA{Zucchini16}. The regime--switching model assumes that a discrete unobservable Markov process switches among a finite set of regimes randomly and that each regime is defined by a particular parameter set. The model is a good fit for some financial data and becomes popular in financial modeling including equity options, bond prices, and others.

The Kalman filtering, which was introduced by \citeA{Kalman60} is an algorithm that provides estimates of some observed and unobserved (state) processes. The Kalman filtering has been demonstrating its usefulness in various applications. It has been used extensively in economics, system theory, the physical sciences, and engineering. In econometrics, the state--space model is usually defined by (i) the observed vector is described in terms of the state vector in linear form (measurement equation), and (ii) the state vector is governed by VAR(1) process (transition equation). To estimate the parameters of the state--space model and to make inferences about the state--space model (smoothing and forecasting), the Kalman filtering can be used, see \citeA{Hamilton94} and \citeA{Lutkepohl05}.

By the CAPM, the required rate of return is modeled by the risk--free rate, beta, and market return. However, the CAPM is sensitive to its inputs. Recently, \citeA{Battulga22a} introduced a stochastic DDM that models the dividends by a compound non--homogeneous Poisson--process and obtained ML estimators and confidence bands of the model's parameters, including the required rate of return. In this paper, instead of the traditional CAPM and its descendant versions, we introduce new estimation methods, which cover the ML methods with regime--switching, the Bayesian method, and the Kalman filtering to estimate the required rate of return on equity.

The rest of the paper is organized as follows: In Section 2, to estimate the required rate of returns on equity for public companies, we introduce the ML method with regime--switching and the Bayesian method. Also, we provide a simple method that evaluates market values of liabilities and portfolio choice theory. Section 3 is devoted to parameter estimation methods for private companies, where we consider the ML method with regime--switching, the Bayesian method, and the Kalman filtering. In Section 4, for selected public companies, we provide numerical results based on our methods. Finally, Section 5 concludes the study.

\section{Parameter Estimation of Public Company}

In this paper, we assume that there are $n$ companies and the companies will not default in the future. As mentioned before the basic idea of all DDMs is that the market price of a stock equals the sum of the stock's next period price and dividend discounted at the required rate of return. Therefore, for successive prices of $i$--th company, the following relation holds 
\begin{equation}\label{02001}
P_{i,t}=(1+k_{i,t}^e)P_{i,t-1}-d_{i,t},~~~i=1,\dots,n ~\text{and}~t=1,2,\dots,
\end{equation}
where $k_{i,t}^e$ is the required rate of return on equity, $P_{i,t}$ is the equity price, and $d_{i,t}$ is the dividend, respectively, at time $t$ of $i$--th company. In this paper, we suppose that the required rate of returns are random variables. For the above DDM equation, if the required rate of return is less than $-1$, namely, $k_{i,t}<-1$, then the sum of the price and dividend, respectively, at time $t$ of $i$--th company takes a negative value, which is an undesirable result. For this reason, we need to write the above DDM equation in the following form
\begin{equation}\label{02002}
P_{i,t}=\exp\{\tilde{k}_{i,t}^e\}P_{i,t-1}-d_{i,t},~~~i=1,\dots,n ~\text{and}~t=1,2,\dots,
\end{equation}
where $\tilde{k}_{i,t}^e:=\ln(1+k_{i,t}^e)$ is a log required rate of return on equity at time $t$ of $i$--th company. To keep notations simple, let $\tilde{k}_t^e:=(\tilde{k}_{1,t}^e,\dots,\tilde{k}_{n,t}^e)'$ be an $(n\times 1)$ required rate of return vector on equity at time $t$, $P_t:=(P_{1,t},\dots,P_{n,t})'$ be an $(n\times 1)$ price vector at time $t$, and $d_t:=(d_{1,t},\dots,d_{n,t})'$ be an $(n\times 1)$ dividend vector at time $t$ of the companies. Then, equation \eqref{02002} can be written in vector form
\begin{equation}\label{02003}
P_t=\exp\{\tilde{k}_t^e\}\odot P_{t-1}-d_t,~~~t=1,2,\dots,
\end{equation}
where $\odot$ is the Hadamard's element--wise product of two vectors. It follows from equation \eqref{02003} that the log required rate of return at time $t$ is represented by
\begin{equation}\label{02004} 
\tilde{k}_t^e=\ln\big((P_t+d_t)\oslash P_{t-1}\big), ~~~t=1,2,\dots,
\end{equation}
where $\oslash$ is the element--wise division of two vectors. It is worth mentioning that because the price vector and dividend vector are known $t$, the value of the log required rate of return vector on equity $k_t^e$ is known at time $t$.

We assume that each of the $n$ companies is financed by identically $m$ different type liabilities. Let $L_{i,j,t}$ and $r_{i,j,t}$ be principal outstanding and payment, including interest payment of $j$--th type liabilities at time $t$ of $i$--th company. The principal outstanding $L_{i,j,t}$ represents the remaining liability immediately after $r_{i,j,t}$ has been paid. It equals the previous period's principal outstanding of the liability, accumulated for one period, minus $r_{i,t}$. Therefore, we have
\begin{equation}\label{02005}
L_{i,j,t}=(1+\bar{k}_{j,t-1})L_{i,j,t-1}-r_{i,j,t},~~~i=1,\dots,n,~j=1,\dots,m,~t=1,2,\dots,
\end{equation}
where $\bar{k}_{j,t-1}$ is an interest rate of the the $j$--th type liability. It should be noted that the interest rate known at time $t-1$. Consequently, the sum $L_{i,j,t}+r_{i,j,t}$ is also known at time $t-1$. If we sum equation \eqref{02005} over all values of $j$, then we obtain
\begin{equation}\label{02006}
L_{i,t}:=\sum_{j=1}^mL_{i,j,t}=\sum_{j=1}^m(1+\bar{k}_{j,t-1})L_{i,j,t-1}-\sum_{j=1}^mr_{i,j,t}=(1+k_{i,t-1})L_{i,t-1}-r_{i,t},
\end{equation}
where $L_{i,t}$ is a total liability (book value) at time $t$, $r_{i,t}$ is total interest payment minus net new borrowing ($L_{i,t}-L_{i,t-1}$) at time $t$, and 
\begin{equation}\label{02007}
k_{i,t-1}=\sum_{j=1}^m\bar{k}_{j,t-1} w_{i,j,t-1}
\end{equation}
with $w_{i,j,t-1}:=\frac{L_{i,j,t-1}}{L_{i,t-1}}$ is a weighted interest rate at time $t-1$ of $i$--th company, respectively. From equation \eqref{02006}, one finds that
\begin{equation}\label{02008}
L_{i,t}=\frac{L_{i,t+1}+r_{i,t+1}}{1+\bar{k}_{i,t}}.
\end{equation}
As a result, if we replace the weighted interest rate $\bar{k}_{i,t}$ in the above equation \eqref{02008} by a weighted market interest rate, then the market value at time $t$ of the $i$--th company's liabilities is obtained by
\begin{equation}\label{02009}
L_{i,t}^m=\frac{L_{i,t+1}+r_{i,t+1}}{1+k_{i,t+1}^\ell}=\frac{I_{i,t}+L_{i,t}}{1+k_{i,t+1}^\ell},
\end{equation}
where $k_{i,t+1}^\ell$ is a weighted market interest rate (required rate of return on debtholders) at time $t+1$ of the liabilities and $I_{i,t}:=k_{i,t}L_{i,t}$ is the total interest payment at time $t$ of the $i$--th company. The weighted market interest rate at time $t+1$ of the liabilities of the $i$--th company is calculated by
\begin{equation}\label{02010}
k_{i,t+1}^\ell=\sum_{j=1}^m\bar{k}_{j,t+1}^\ell w_{i,j,t},
\end{equation}
where $\bar{k}_{j,t+1}^\ell$ is market interest rate at time $t+1$ of the $j$--th type liability. The formula of the market value of the liabilities, given in equation \eqref{02008} also holds for individual liabilities, namely,
\begin{equation}\label{02011}
L_{i,j,t}^m=\frac{I_{i,j,t}+L_{i,j,t}}{1+\bar{k}_{j,t+1}^\ell}, ~~~j=1,\dots,m,
\end{equation}
where $I_{i,j,t}:=\bar{k}_{j,t}L_{i,j,t}$ is the interest payment at time $t$ for $j$--th type liability of the $i$--th company. It can be shown that similarly to equation \eqref{02001}, for successive market values of the $i$--th company's liabilities, we have
\begin{equation}\label{02012}
L_{i,t}^m=(1+k_{i,t}^\ell)L_{i,t-1}^m-r_{i,t},~~~t=1,2,\dots.
\end{equation}
Consequently, if a company is financed by liabilities, which are publicly traded in the exchanges, then one can estimate the required rate of return on debtholders using methods, which will appear in this Section, see below.

We assume that the log required rate of return vector at time $t$ on equities, $\tilde{k}_t^e$, places first $n$ components  of a $(n+\ell)$ Markov--Switching Vector Autoregressive (MS--VAR($p$)) process with order $p$ and regimes $N$. Let us denote the dimension of the MS--VAR$(p)$ process by $\tilde{n}$, i.e., $\tilde{n}:=n+\ell$. As the log required rate of returns on stocks depends on macroeconomic variables and firm--specific variables, such as GDP, inflation, key financial ratios of the companies, and so on, the last $\ell$ components of the MS--VAR$(p)$ process $y_t$ correspond to the economic variables that affect the log required rate of returns on equities of the companies. The economic variables may or may not contain dividends. The MS--VAR($p$) process $y_t$ is given by the following equation
\begin{equation}\label{02013}
y_t=A_0(s_t)\psi_t+A_1(s_t)y_{t-1}+\dots+A_p(s_t)y_{t-p}+\xi_t,
\end{equation}
where $y_t=(y_{1,t},\dots,y_{\tilde{n},t})'$ is an $(\tilde{n}\times 1)$ random vector, $\psi_t=(\psi_{1,t},\dots,\psi_{l,t})'$ is a $(l\times 1)$ random vector of exogenous variables, $\xi_t=(\xi_{1,t},\dots,\xi_{\tilde{n},t})'$ is an $(\tilde{n}\times 1)$ residual process, $s_t$ is an unobserved regime at time $t$, which is governed by a Markov chain with $N$ states, $A_0(s_t)$ is an $(\tilde{n}\times l)$ coefficient matrix at regime $s_t$ that corresponds to the vector of exogenous variables, for $i=1,\dots,p$, $A_i(s_t)$ are $(\tilde{n}\times \tilde{n})$ coefficient matrices at regime $s_t$ that correspond to $y_{t-1},\dots,y_{t-p}$. 

For the residual process $\xi_t$, we assume that it has $\xi_t:=\Sigma_t^{1/2}(\bar{s}_t)\varepsilon_t$ representation, see \citeA{Lutkepohl05} and \citeA{McNeil05}, where $\bar{s}_t=(s_1,\dots,s_t)'$ is a $(t\times 1)$ vector of up to and including time $t$ regimes, $\Sigma_t^{1/2}(\bar{s}_t)$ is Cholesky factor of a $(\tilde{n}\times \tilde{n})$ positive definite matrix $\Sigma_t(\bar{s}_t)$, which is measurable with respect to $\sigma$--field $\{\mathcal{F}_{t-1},\bar{s}_t\}$ and depends on coefficient matrix $\Gamma(s_t):=[B_0(s_t):B_1(s_t):\dots:B_{p_*+q_*}(s_t)]$. Here $\mathcal{F}_t$ is a $\sigma$--field, defined below, and $B_0(s_t)$ is an $(n_*\times l_*)$ matrix, for $i=1,\dots,p_*+q_*$, $B_i(s_t)$ are $(n_*\times n_*)$ matrices, and $\varepsilon_1,\dots,\varepsilon_T$ is a random sequence of independent identically multivariate normally distributed random vectors with means of 0 and covariance matrices of $n$ dimensional identity matrix $I_n$. Then, in particular, for multivariate GARCH process of $(p_*,q_*)$ order, dependence of $\Sigma_t^{1/2}$ on $\Gamma(s_t)$ is given by 
\begin{equation}\label{02014}
\text{vech}\big(\Sigma_t(\bar{s}_t)\big)=B_0(s_t)+\sum_{i=1}^{p_*}B_i(s_t)\text{vech}\big(\xi_{t-i}\xi_{t-i}'\big)+\sum_{j=1}^{q_*}B_{m_1+j,t}(s_t)\text{vech}(\Sigma_{t-j}(\bar{s}_{t-j})),
\end{equation}
where $B_0(s_t)\in \mathbb{R}^{n(n+1)/2}$ and $B_i(s_t)\in \mathbb{R}^{[n(n+1)/2]\times [n(n+1)/2]}$ for $i=1,\dots, p_*+q_*$ are suitable random vector and matrices and the vech is an operator that stacks elements on and below a main diagonal of
a square matrix. If we assume that in addition to an initial information $\mathcal{F}_0:=\{y_{1-p},\dots,y_0,\psi_1,\dots,\psi_T,\Sigma_{1-q_*},\dots,\Sigma_0\}$, there are $T$ observations of the MS--VAR($p$) process $y_t$, then equation \eqref{02013} can be compactly written by
\begin{equation}\label{02015}
y_t=\Pi(s_t)\mathsf{Y}_{t-1}+\xi_t,~~~t=1,\dots,T,
\end{equation}
where $\Pi(s_t):=[A_0(s_t): A_1(s_t):\dots:A_p(s_t)]$ is a $(\tilde{n}\times[l+\tilde{n}p])$ coefficient matrix at regime $s_t$, which consist of all the coefficient matrices and $\mathsf{Y}_{t-1}:=(\psi_t',y_{t-1}',\dots,y_{t-p}')'$ is an $([l+\tilde{n}p]\times 1)$ vector, which consist of exogenous variable $\psi_t$ and last $p$ lagged values of the process $y_t$. 

Let for each regime $j=1,\dots,N$, $\pi(j):=\text{vec}(\Pi(j))$ is an $\big(\tilde{n}(l+\tilde{n}p)\times 1\big)$ vector, corresponding to the matrix $\Pi(j)$ and $\gamma(j):=\text{vec}(\Gamma(j))$ is $\big([n_*(l_*+n_*(p_*+q_*))]\times 1\big)$ vector, corresponding to the matrix $\Gamma(j)$, where for a generic $(n\times m)$ matrix $A$, $\mathrm{vec}(A)$ is an operator that transform $A$ into $(nm\times 1)$ vector by stacking columns. For our model, the coefficient vector is $\big(\pi(1)',\gamma(1)'\big)'$ when the process is in regime 1, $\big(\pi(2)',\gamma(2)'\big)'$ when the process is in regime 2, and so on. 

Since we assume that the regime--switching process $s_t$ is governed by first--order homogeneous Markov chain, a conditional probability that the regime at time $t$, $s_t$ equals some particular value conditional on the past regimes, $s_{t-1},s_{t-2},\dots,s_1$ depends only through the most recent regime at time $t-1$, $s_{t-1}$, and does not depend on time, that is, 
\begin{equation}\label{02016}
p_{ij}:=\mathbb{P}(s_t=j|s_{t-1}=i)=\mathbb{P}(s_t=j|s_{t-1}=i,s_{t-2}=s_{t-2},\dots,s_1=s_1),~~~i,j=1,\dots,N.
\end{equation}
If we collect all the conditional probabilities $p_{ij}$ into a matrix $\mathsf{P}$, then we obtain a transition probability matrix of the regime--switching process $s_t$
\begin{equation}\label{02017}
\mathsf{P}=\begin{bmatrix}
p_{11} & p_{12} & \dots & p_{1N}\\
p_{21} & p_{22} & \dots & p_{2N}\\
\vdots & \vdots & \ddots & \vdots\\
p_{N1} & p_{N2} & \dots & p_{NN}\\
\end{bmatrix}.
\end{equation}
Observe that sums of all rows of the transition probability matrix $\mathsf{P}$ equal 1, that is, for all $i=1,\dots,N$, $p_{i1}+\dots+p_{iN}=1.$ 

\subsection{Regime Switching Estimation}\label{sub01}

This Subsection is devoted to regime--switching estimators of parameters of the required rate of return om equity and is based on the book of \citeA{Hamilton94}. For $t=0,\dots,T$, let us denote available information at time $t$ by $\mathcal{F}_t$, which consists of the required rate of returns on equities, economic variables, and exogenous variables: $\mathcal{F}_t:=(\mathcal{F}_0,y_1,\dots,y_t)'.$ Then, it is clear that the log--likelihood function of our model is given by the following equation
\begin{equation}\label{02018}
\mathcal{L}(\theta)=\sum_{t=1}^T\ln\big(f(y_t|\mathcal{F}_{t-1};\theta)\big)
\end{equation}
where $\theta:=\big(\pi(1)',\dots,\pi(N)',\gamma(1)',\dots,\gamma(N)',\rho',\text{vec}(\mathsf{P})'\big)'$ is a vector, which consists of all population parameters of the model and $f(y_t|\mathcal{F}_{t-1};\theta)$ is a conditional density function of the random vector $y_t$ given the information $\mathcal{F}_{t-1}$. Here $\rho:=(\mathbb{P}(s_1|\mathcal{F}_0),\dots,\mathbb{P}(s_N|\mathcal{F}_0))'$ is an $(N\times 1)$ initial probability vector. The log--likelihood function is used to obtain the maximum likelihood estimator of the parameter vector $\theta$. Note that the log--likelihood function depends on all observations, which are collected in $\mathcal{F}_T$, but does not depend on regime--switching process $s_t$, whose values are unobserved. If we assume that the regime--switching process in regime $j$ at time $t$, then because conditional on the information $\mathcal{F}_{t-1}$, $\xi_t$ follows a multivariate normal distribution with mean zero and covariance matrix $\Sigma_t(j)$, the conditional density function of the random vector $y_t$ is given by the following equation
\begin{eqnarray}\label{02019}
\eta_{tj}&:=&f(y_t|s_t=j,\mathcal{F}_{t-1};\alpha)\\
&=&\frac{1}{(2\pi)^{\tilde{n}/2}|\Sigma_t(j)|^{1/2}}\exp\bigg\{-\frac{1}{2}\Big(y_t-\Pi(j)\mathsf{Y}_{t-1}\Big)'\Sigma_t^{-1}(j)\Big(y_t-\Pi(j)\mathsf{Y}_{t-1}\Big)\bigg\}\nonumber
\end{eqnarray}
for $t=1,\dots,T$ and $j=1,\dots,N$, where $\alpha:=\big(\pi(1)',\dots,\pi(N)',\gamma(1)',\dots,\gamma(N)'\big)'$ is a parameter vector, which differs from the vector of all parameters $\theta$ by the initial probability vector $\rho$ and transition probability matrix $\mathsf{P}$. As a result, since $\Pi(j)\mathsf{Y}_{t-1}=\big(\mathsf{Y}_{t-1}'\otimes I_{\tilde{n}}\big)\pi(j)$, a log of the conditional density function $\eta_{tj}$ is represented by
\begin{eqnarray}\label{02020}
\ln(\eta_{tj})&=&-\frac{\tilde{n}}{2}\ln(2\pi)-\frac{1}{2}\ln(|\Sigma_t(j)|)\\
&-&\frac{1}{2}\Big(y_t-\big(\mathsf{Y}_{t-1}'\otimes I_{\tilde{n}}\big)\pi(j)\Big)'\Sigma_t^{-1}(j)\Big(y_t-\big(\mathsf{Y}_{t-1}'\otimes I_{\tilde{n}}\big)\pi(j)\Big),\nonumber
\end{eqnarray}
where $\otimes$ is the Kronecker product of two matrices.

For all $t=1,\dots,T$, we collect the conditional density functions of the price at time $t$ into an $(n\times 1)$ vector $\eta_t$, that is, $\eta_t:=(\eta_{t1},\dots,\eta_{tN})'$. Let us denote a probabilistic inference about the value of the regime--switching process $s_t$ equals to $j$, based on the information $\mathcal{F}_t$ and the parameter vector $\theta$ by $\mathbb{P}(s_t=j|\mathcal{F}_t,\theta)$. Collect these conditional probabilities $\mathbb{P}(s_t=j|\mathcal{F}_t,\theta)$ for $j=1,\dots,N$ into an $(N\times 1)$ vector $z_{t|t}$, that is, $z_{t|t}:=\big(\mathbb{P}(s_t=1|\mathcal{F}_t;\theta),\dots,\mathbb{P}(s_t=N|\mathcal{F}_t;\theta)\big)'$. Also, we need a probabilistic forecast about the value of the regime--switching process at time $t+1$ equals $j$ conditional on data up to and including time $t$. Collect these forecasts into an $(N\times 1)$ vector $z_{t+1|t}$, that is, $z_{t+1|t}:=\big(\mathbb{P}(s_{t+1}=1|\mathcal{F}_t;\theta),\dots,\mathbb{P}(s_{t+1}=N|\mathcal{F}_t;\theta)\big)'$.  

The probabilistic inference and forecast for each time $t=1,\dots,T$ can be found by iterating on the following pair of equations: 
\begin{equation}\label{02021}
z_{t|t}=\frac{(z_{t|t-1}\odot\eta_t)}{i_N'(z_{t|t-1}\odot\eta_t)}~~~\text{and}~~~z_{t+1|t}=\mathsf{P}'z_{t|t},~~~t=1,\dots,T,
\end{equation}
where $\eta_t$ is the $(N\times 1)$ vector, whose $j$-th element is given by equation \eqref{02020}, $\mathsf{P}$ is the $(N\times N)$ transition probability matrix, which is given by equation \eqref{02017}, and $i_N$ is an $(N\times 1)$ vector, whose elements equal 1. Given a starting value $\rho=z_{1|0}$ and an assumed value for the population parameter vector $\theta$, one can iterate on \eqref{02021} for $t=1,\dots,T$ to calculate the values of $z_{t|t}$ and $z_{t+1|t}$. To obtain MLE of the population parameters, in addition to the inferences and forecasts we need a smoothed inference about the regime--switching process was in at time $t$ based on full information $\mathcal{F}_T$. Collect these smoothed inferences into an $(N\times 1)$ vector $z_{t|T}$, that is, $z_{t|T}:=\big(\mathbb{P}(s_t=1|\mathcal{F}_T;\theta),\dots,\mathbb{P}(s_t=N|\mathcal{F}_T;\theta)\big)'$. The smoothed inferences can be obtained by using the \citeauthor{Kim94}'s \citeyear{Kim94} smoothing algorithm:
\begin{equation}\label{02022}
z_{t|T}=z_{t|t}\odot\big\{\mathsf{P}'(z_{t+1|T}\oslash z_{t+1|t})\big\},~~~t=T-1,\dots,1,
\end{equation}
where $\oslash$ is an element--wise division of two vectors. The smoothed probabilities $z_{t|T}$ are found by iterating on \eqref{02022} backward for $t=T-1,\dots,1$. This iteration is started with $z_{T|T}$, which is obtained from \eqref{02021} for $t=T$.

If the initial probability $\rho$ does not depend on the other parameters, then according to \citeA{Hamilton90}, maximum likelihood estimators of $(i,j)$-th element of the transition probability matrix $\mathsf{P}$, the parameter vector $\alpha$ that governs the conditional density functions \eqref{02019}, and the initial probability $\rho$ are obtained from the following systems of equations
\begin{eqnarray}
\hat{p}_{ij}&=&\frac{\sum_{t=2}^T\mathbb{P}\big(s_{t-1}=i,s_t=j|\mathcal{F}_T;\hat{\theta}\big)}{\sum_{t=2}^T(z_{t-1|T})_i},\label{02023}\\
0&=&\sum_{t=1}^T\bigg(\frac{\partial \ln(\eta_t)}{\partial\alpha'}\bigg)'z_{t|T},\label{02024}\\
\hat{\rho}&=&z_{1|T},\label{02025}
\end{eqnarray}
where ${\partial \ln(\eta_t)}/{\partial\alpha'}$ is an $\big(N \times [\tilde{n}(l+\tilde{n}p)+n_*(l_*+n_*(p_*+q_*))]\big)$ matrix of derivatives of the logs of the conditional densities and due to the Kim's smoothing algorithm, the numerator of equation \eqref{02023} can be calculated by
\begin{equation}\label{02026}
\mathbb{P}\big(s_{t-1}=i,s_t=j|\mathcal{F}_T;\theta\big)=p_{ij}(z_{t|T})_j(z_{t-1|t-1})_i/{(z_{t|t-1})_j}.
\end{equation}

To simplify notations for MLE that correspond to the parameter vector $\alpha$, for each regime $j=1,\dots,N$, let $\bar{\mathsf{Y}}_j:=\big[\bar{\mathsf{Y}}_{0,j}:\dots:\bar{\mathsf{Y}}_{T-1,j}\big]$ be an $\big([l+\tilde{n}p]\times T\big)$ matrix, which is adjusted by the regime $j$ and whose $t$--th column is given by an $\big([l+\tilde{n}p]\times 1\big)$ vector $\bar{\mathsf{Y}}_{t-1,j}:=\mathsf{Y}_{t-1}\sqrt{(z_{t|T})_j}$ and $\bar{y}_j:=\big[\bar{y}_{1,j}:\dots:\bar{y}_{T,j}\big]$ be a $\big(\tilde{n}\times T\big)$ matrix, which is adjusted by the regime $j$ and whose $t$--th column is given by a $\big(\tilde{n}\times 1\big)$ vector $\bar{y}_{t,j}:=y_t\sqrt{(z_{t|T})_j}$. 

Firstly, let us assume that for each $j=1,\dots,N$, the covariance matrix at regime $j$ is homoscedastic. Then, according to equation \eqref{02020}, partial derivatives of the log conditional density function $\ln(\eta_{tj})$ with respect to the vectors $\pi(m)$, $m=1,\dots,N$ is given by
\begin{equation}\label{02027}
\frac{\partial \ln(\eta_{tj})}{\partial \pi(m)'}=
\begin{cases}
\Big(y_t-\big(\mathsf{Y}_{t-1}'\otimes I_{\tilde{n}}\big)\pi(j)\Big)'\Sigma^{-1}(j)\big(\mathsf{Y}_{t-1}'\otimes I_{\tilde{n}}\big) & \text{for}~~~j=m,\\
0 & \text{for}~~~j\neq m.
\end{cases}
\end{equation}
Thus, due to equation \eqref{02024}, one gets that
\begin{equation}\label{02028}
\sum_{t=1}^T\Big(\bar{y}_{t,j}-\big(\bar{\mathsf{Y}}_{t-1,j}'\otimes I_{\tilde{n}}\big)\pi(j)\Big)'\Sigma^{-1}(j)\big(\bar{\mathsf{Y}}_{t-1,j}'\otimes I_{\tilde{n}}\big)=0
\end{equation}
for $j=1,\dots,N$. Consequently, for each regime $j=1,\dots,N$, ML estimator of the parameter vector $\pi(j)$ is obtained by
\begin{equation}\label{02029}
\hat{\pi}(j):=\Bigg(\sum_{t=1}^T\big(\bar{\mathsf{Y}}_{t-1,j}\otimes I_{\tilde{n}}\big)\Sigma^{-1}(j)\big(\bar{\mathsf{Y}}_{t-1,j}'\otimes I_{\tilde{n}}\big)\Bigg)^{-1}\sum_{t=1}^T\big(\bar{\mathsf{Y}}_{t-1,j}\otimes I_{\tilde{n}}\big)\Sigma^{-1}(j)\bar{y}_{t,j}.
\end{equation}
Since $\big(\bar{\mathsf{Y}}_{t-1,j}\otimes I_{\tilde{n}}\big)\Sigma^{-1}(j)=\big(\bar{\mathsf{Y}}_{t-1,j}\otimes\Sigma^{-1}(j)\big)$, we find that
\begin{equation}\label{02030}
\sum_{t=1}^T\big(\bar{\mathsf{Y}}_{t-1,j}\otimes I_{\tilde{n}}\big)\Sigma^{-1}(j)\big(\bar{\mathsf{Y}}_{t-1,j}'\otimes I_{\tilde{n}}\big)=\Big(\bar{\mathsf{Y}}_{j}\bar{\mathsf{Y}}_{j}'\otimes \Sigma^{-1}(j)\Big)
\end{equation}
and
\begin{equation}\label{02031}
\sum_{t=1}^T\big(\bar{\mathsf{Y}}_{t-1,j}\otimes I_{\tilde{n}}\big)\Sigma^{-1}(j)\bar{y}_{t,j}=\Big(\bar{\mathsf{Y}}_j\otimes \Sigma^{-1}(j)\Big)\text{vec}(\bar{y}_j).
\end{equation}
Therefore, the ML estimator $\hat{\pi}(j)$ is represented by
\begin{equation}\label{02032}
\hat{\pi}(j)=\text{vec}\big(\hat{\Pi}(j)\big)=\Big(\big(\bar{\mathsf{Y}}_{j}\bar{\mathsf{Y}}_{j}'\big)^{-1}\bar{\mathsf{Y}}_j\otimes I_{\tilde{n}}\Big)\text{vec}(\bar{y}_j)=\text{vec}\Big(\bar{y}_j\bar{\mathsf{Y}}_j'\big(\bar{\mathsf{Y}}_{j}\bar{\mathsf{Y}}_{j}'\big)^{-1}\Big).
\end{equation}
As a result, for each regime $j=1,\dots,N$, ML estimator of the parameter $\Pi(j)$ is given by the following equation
\begin{equation}\label{02033}
\hat{\Pi}(j)=\bar{y}_j\bar{\mathsf{Y}}_j'\big(\bar{\mathsf{Y}}_{j}\bar{\mathsf{Y}}_{j}'\big)^{-1}, ~~~j=1,\dots,N.
\end{equation}
On the other hand, due to equation \eqref{02020}, we have
\begin{equation}\label{02034}
\frac{\partial \ln(\eta_{tj})}{\partial \Sigma(m)}=
\begin{cases}
\begin{matrix}\displaystyle-\frac{1}{2}\Sigma_t^{-1}(j)+\frac{1}{2}\Sigma_t^{-1}(j)\Big(y_t-\big(\mathsf{Y}_{t-1}'\otimes I_{\tilde{n}}\big)\pi(j)\Big)\\
\times\Big(y_t-\big(\mathsf{Y}_{t-1}'\otimes I_{\tilde{n}}\big)\pi(j)\Big)'\Sigma_t^{-1}(j)\end{matrix} & \text{for}~~~j=m,\\
0 & \text{for}~~~j\neq m.
\end{cases}
\end{equation}
Consequently, by equation \eqref{02024} ML estimator of the parameter $\Sigma(j)$ is obtained by
\begin{equation}\label{02035}
\hat{\Sigma}(j)=\frac{1}{\sum_{t=1}^T(z_{t|T})_j}\sum_{t=1}^T\Big(\bar{y}_{t,j}-\hat{\Pi}(j)\mathsf{Y}_{t-1,j}\Big)\Big(\bar{y}_{t,j}-\hat{\Pi}(j)\mathsf{Y}_{t-1,j}\Big)'
\end{equation}
for $j=1,\dots,N$,

Secondly, we suppose that for each $j=1,\dots,N$, the covariance matrix is homoscedastic and does not depend on regimes, $\Sigma_t(j)=\Sigma$. Then, similarly to before, it can be shown that maximum likelihood estimators of the parameters $\Pi(j)$ and $\Sigma$ are obtained by
\begin{equation}\label{02036}
\hat{\Pi}(j)=\bar{y}_j\bar{\mathsf{Y}}_j'(\bar{\mathsf{Y}}_j\bar{\mathsf{Y}}_j')^{-1}
\end{equation}
for $j=1,\dots,N$ and
\begin{equation}\label{02037}
\hat{\Sigma}=\frac{1}{T}\sum_{t=1}^T\sum_{j=1}^N\Big(\bar{y}_{t,j}-\hat{\Pi}(j)\mathsf{Y}_{t-1,j}\Big)\Big(\bar{y}_{t,j}-\hat{\Pi}(j)\mathsf{Y}_{t-1,j}\Big)'.
\end{equation}

Thirdly, we assume that there is one regime ($N=1$) and the covariance matrix is homoscedastic, $\Sigma_t(j)=\Sigma$ and $\Pi(j)=\Pi$. Then, as before, maximum likelihood estimators of the parameters $\Pi$ and $\Sigma$ are found by
\begin{equation}\label{02038}
\hat{\Pi}=\bar{y}\bar{\mathsf{Y}}'(\bar{\mathsf{Y}}\bar{\mathsf{Y}}')^{-1}
\end{equation}
and
\begin{equation}\label{02039}
\hat{\Sigma}=\frac{1}{T}\sum_{t=1}^T\Big(y_{t}-\hat{\Pi}\mathsf{Y}_{t-1}\Big)\Big(y_{t}-\hat{\Pi}\mathsf{Y}_{t-1}\Big)',
\end{equation}
where $\bar{\mathsf{Y}}:=\big[\bar{\mathsf{Y}}_{0}:\dots:\bar{\mathsf{Y}}_{T-1}\big]$ and $\bar{y}:=\big[\bar{y}_{1}:\dots:\bar{y}_{T}\big]$.

Fourthly, we assume that there is one regime ($N=1$), one company $(n=1)$, no exogenous variables except 1, and no economic variables, order of AR process equals 0, and a variance of the white noise process $\xi_t$ is homoscedastic, $\text{Var}(\xi_t)=\sigma^2$. In this assumption, equation \eqref{02013} becomes AR(0) process
\begin{equation}\label{02148}
\tilde{k}_t=a_0+\xi_t,
\end{equation}
where $\tilde{k}_t$ is the log required rate of return on equity of the company. Then, it follows from equations \eqref{02038} and \eqref{02039}, maximum likelihood estimators of the parameters $a_0$ and $\sigma^2$ are obtained by
\begin{equation}\label{02149}
\hat{a}_0=\frac{1}{T}\sum_{t=1}^T\tilde{k}_t~~~\text{and}~~~\hat{\sigma}^2=\frac{1}{T}\sum_{t=1}^T(\tilde{k}_t-\hat{a}_0)^2.
\end{equation}
Consequently, the maximum likelihood estimators of the parameters $a_0$ equals geometric average of the required rate of returns
\begin{equation}\label{02150}
\hat{a}_0=\sqrt[T]{(1+k_1)\dots (1+k_T)}
\end{equation}
and $(1-\alpha)100\%$ confidence intervals of the parameters $a_0$ and $\sigma^2$ are 
\begin{equation}\label{02151}
\hat{a}_0-t_{1-\alpha/2}(T-1)\frac{\hat{\sigma}}{\sqrt{T-1}}\leq a_0\leq \hat{a}_0+t_{1-\alpha/2}(T-1)\frac{\hat{\sigma}}{\sqrt{T-1}}
\end{equation}
and
\begin{equation}\label{02152}
\frac{T\hat{\sigma}^2}{\chi_{1-\alpha/2}^2(T-1)}\leq \sigma^2\leq \frac{T\hat{\sigma}^2}{\chi_{\alpha/2}^2(T-1)},
\end{equation}
where $t_{1-\alpha/2}(T-1)$ is a $(1-\alpha/2)$ quantile of the student $t$ distribution with $(T-1)$ degrees of freedom and $\chi_{\alpha/2}^2(T-1)$ is a $\alpha/2$ quantile of the chi--square distribution with $(T-1)$ degrees of freedom. 

From equation \eqref{02148}, a point prediction of the log required rate of return on equity equals $\tilde{k}=\hat{a}_0$. Let us assume that true value of the prediction is $\tilde{k}_0=a_0+\xi_0$. Then, a prediction error equals $e_0:=\tilde{k}_0-\tilde{k}=\xi_0$. Then, it is clear that $e_0/\sigma\sim \mathcal{N}(0,1)$. The ML estimator of the parameter $\sigma^2$ can be written by
\begin{equation}\label{02153}
\hat{\sigma}^2=\frac{1}{T}\sum_{t=1}^T(\tilde{k}_t-\hat{a}_0)^2=\frac{1}{T}\sum_{t=1}^T(\xi_t-\bar{\xi})^2=\frac{1}{T}\xi'A\xi,
\end{equation}
where $\xi:=(\xi_1,\dots,\xi_T)'$ is a $(T\times 1)$ vector, $\bar{\xi}=\frac{1}{T}\sum_{t=1}^T\xi_t$ is the mean of the vector $\xi$, and $A:=I_T-\frac{1}{T}i_Ti_T'$ is a $(T\times T)$ symmetric idempotent matrix with rank $T-1$. Since $\xi\sim \mathcal{N}(0,\sigma^2I_T)$, it holds $\frac{1}{\sigma}\xi'A\xi\sim \chi^2(T-1)$, see, e.g., \citeA{Johnston97}. Because $\xi_0$ is independent of $\xi$, one finds that
\begin{equation}\label{02154}
\frac{e_0}{\sigma}\Bigg/\sqrt{\frac{\frac{1}{\sigma}\xi'A\xi}{T-1}}=\frac{\tilde{k}_0-\tilde{k}}{\hat{\sigma}}\sqrt{\frac{T-1}{T}}\sim t(T-1).
\end{equation}
Consequently, $(1-\alpha)100\%$ confidence interval for the log required rate of return on equity is given by the following equation
\begin{equation}\label{02155}
\hat{a}_0-t_{1-\alpha/2}(T-1)\sqrt{\frac{T}{T-1}}\hat{\sigma}\leq \tilde{k}_0\leq \hat{a}_0+t_{1-\alpha/2}(T-1)\sqrt{\frac{T}{T-1}}\hat{\sigma}.
\end{equation}
As a result, $(1-\alpha)100\%$ confidence interval for the required rate of return on equity is
\begin{equation}\label{02156}
\exp\bigg\{\hat{a}_0-t_{1-\alpha/2}(T-1)\sqrt{\frac{T}{T-1}}\hat{\sigma}\bigg\}-1\leq k_0\leq \exp\bigg\{\hat{a}_0+t_{1-\alpha/2}(T-1)\sqrt{\frac{T}{T-1}}\hat{\sigma}\bigg\}-1.
\end{equation}
The confidence bands will be used in Section 4.

The maximum likelihood estimator of the parameter vector $\theta$ is obtained by the zig--zag iteration method using equations \eqref{02021}--\eqref{02025}, \eqref{02033}, \eqref{02035}, and \eqref{02037}. 

\subsection{The Bayesian Estimation}\label{sub02}

The VAR$(p)$ process is the workhorse model for empirical macroeconomics. However, if the number of variables in the system increases or the time lag is chosen high, then too many parameters need to be estimated. This will reduce the degrees of freedom of the model and entails a risk of over-parametrization. For this reason, in this subsection, we consider the Bayesian analysis for the VAR$(p)$ process $y_t$. In order to simplify calculations, we assume that our model has one regime, that is, $N=1$. Under the assumption, our model \eqref{02015} is given by
\begin{equation}\label{02041}
y_t=\Pi\mathsf{Y}_{t-1}+\xi_t, ~~~t=1,\dots,T.
\end{equation}
where $y_t$ is the $(\tilde{n}\times 1)$ vector, which includes the log required rate of return vector on equity $\tilde{k}_t$, $\Pi$ is the $(\tilde{n}\times [l+\tilde{n} p])$ random matrix, $\mathsf{Y}_{t-1}:=\big(\psi_t',y_{t-1}',\dots,y_{t-p}'\big)'$ is the $([l+\tilde{n} p]\times 1)$ vector, and conditional on $\Sigma$, $\xi_t$ is the $(\tilde{n}\times 1)$ white noise process with a random covariance matrix $\Sigma=\text{Var}(\xi_t)$. To obtain the Bayesian estimator of the model, we need two representations of the VAR$(p)$ process $y_t$, namely
\begin{itemize}
\item[($i$)] the first one is
\begin{equation}\label{02042}
y=\mathsf{Y}\pi+\xi,
\end{equation}
where $y=(y_1',\dots,y_T')'$ is an $([\tilde{n}T]\times 1)$ random vector, $\mathsf{Y}:=\text{diag}\{\mathsf{Y}_0'\otimes I_{\tilde{n}},\dots,\mathsf{Y}_{T-1}'\otimes I_{\tilde{n}}\}$ is $\big([\tilde{n}T]\times [\tilde{n}(l+\tilde{n} p]\big)$ matrix, $\pi:=\text{vec}(\Pi)$ is an $\big([\tilde{n}(l+\tilde{n} p)]\times 1\big)$ vector, which is a vectorization of the random matrix $\Pi$, and conditional on $\Sigma$, $\xi:=(\xi_1',\dots,\xi_T')'$ is an $(\tilde{n}T\times 1)$ white noise vector and its distribution is $\xi|\Sigma\sim \mathcal{N}(0,I_T\otimes \Sigma)$. From this representation, likelihood function is obtained by
\begin{equation}\label{02043}
f(y|\pi,\Sigma,\mathsf{Y})=\frac{1}{(2\pi)^{\tilde{n}T/2}|\Sigma|^{T/2}}\exp\bigg\{-\frac{1}{2}\Big(y-\mathsf{Y}\pi\Big)'\big(I_T\otimes \Sigma^{-1}\big)\Big(y-\mathsf{Y}\pi\Big)\bigg\}
\end{equation}
\item[($ii$)] and the second one is
\begin{equation}\label{02044}
\bar{y}=\Pi \bar{\mathsf{Y}}+\bar{\xi},
\end{equation}
where $\bar{y}:=[y_1:\dots:y_T]$ is an $(\tilde{n}\times T)$ matrix, $\bar{\mathsf{Y}}:=[\mathsf{Y}_0:\dots:\mathsf{Y}_T]$ is an $([l+\tilde{n} p]\times T)$ matrix, and $\bar{\xi}:=[\xi_1:\dots:\xi_T]$ is an $(\tilde{n}\times T)$ white noise matrix. It is the well--known fact that for suitable matrices $A,B,C,D$,
\begin{equation}\label{02045}
\text{vec}(A)'(B\otimes C)\text{vec}(D)=\text{tr}(DB'A'C).
\end{equation}
As a result, the likelihood function can be written by
\begin{equation}\label{02046}
f(y|\Pi,\Sigma,\bar{\mathsf{Y}})=\frac{1}{(2\pi)^{\tilde{n}T/2}|\Sigma|^{T/2}}\exp\bigg\{-\frac{1}{2}\text{tr}\Big(\big(\bar{y}-\Pi\bar{\mathsf{Y}}\big)\big(\bar{y}-\Pi\bar{\mathsf{Y}}\big)'\Sigma^{-1}\Big)\bigg\}.
\end{equation}
\end{itemize}

In the Bayesian analysis, it assumes that an analyst has a prior probability belief $f(\theta)$ about the unknown parameter $\theta:=(\Pi,\Sigma)$, where $f(\theta)$ is a prior density function of the parameter $\theta$. Let us assume that prior density functions of the parameters $\pi$ and $\Sigma$ are multivariate normal with mean $\pi_0$ and covariance matrix $(\Sigma\otimes \Lambda_0)$ conditional on $\Sigma$ and inverse--Wishart distribution with shape parameters $\nu_0$ and scale matrix $V_0$, respectively, where $\Lambda_0$ is an $([l+\tilde{n}p]\times[l+\tilde{n}p])$ matrix, $\nu_0$ is a real number such that $\nu_0>\tilde{n}-1$, and $V_0$ is an $(\tilde{n}\times \tilde{n})$ positive definite matrix. Thus, due to equation \eqref{02045}, the prior density functions are proportional to
\begin{equation}\label{02047}
f(\Sigma|\nu_0,V_0)\propto |\Sigma|^{-(\nu_0+\tilde{n}+1)/2}\exp\bigg\{-\frac{1}{2}\text{tr}\Big(V_0\Sigma^{-1}\Big)\bigg\}
\end{equation}
and
\begin{eqnarray}\label{02048}
f(\pi|\Sigma,\pi_0,\Lambda_0)&\propto& |\Sigma|^{-(l+\tilde{n} p)/2}\exp\bigg\{-\frac{1}{2}\Big(\big(\pi-\pi_0\big)'\big(\Sigma^{-1}\otimes\Lambda_0^{-1}\big)\big(\pi-\pi_0\big)\Big)\bigg\}\\
&=&|\Sigma|^{-(l+\tilde{n} p)/2}\exp\bigg\{-\frac{1}{2}\text{tr}\Big(\big(\Pi-\Pi_0\big)\Lambda_0^{-1}\big(\Pi-\Pi_0\big)'\Sigma^{-1}\Big)\bigg\},\nonumber
\end{eqnarray}
where $\propto$ is the notation of proportionality and $\Pi_0$ is an $(\tilde{n}\times [l+\tilde{n}p])$ known matrix, which satisfy $\pi_0=\text{vec}(\Pi_0)$. From the conditional density function in equation \eqref{02048}, one can deduce that the analyst's best guess of the parameter $\pi$ is the vector $\pi_0$, and the confidence in this guess is summarized by the matrix $(\Sigma\otimes \Lambda_0)$ and less confidence is represented by larger diagonal elements of $\Lambda_0$. 

After values of $y$ and $\mathsf{Y}$ is observed, the likelihood function $f(y|\Pi,\Sigma,\mathsf{Y})$ will update our beliefs about the parameter $(\Pi,\Sigma)$. Which leads to a posterior density function $f(\Pi,\Sigma|y,\mathsf{Y})$. For each numerical value of the parameter $(\Pi,\Sigma)$, the posterior density $f(\Pi,\Sigma|y,\mathsf{Y})$ describes our belief that $(\Pi,\Sigma)$ is the true value, having observed values of $y$ and $\mathsf{Y}$. It follows from equations \eqref{02046}--\eqref{02048} that a posterior density of the parameter $(\Pi,\Sigma)$ is given by
\begin{eqnarray}\label{02049}
f(\Pi,\Sigma|\bar{y},\bar{\mathsf{Y}})&\propto& f(\Pi|\Sigma,\pi_0,\Lambda_0)f(\Sigma|\nu_0,V_0)f(y|\Pi,\Sigma,\mathsf{Y})\nonumber\\
&\propto& |\Sigma|^{-(\nu_0+l+\tilde{n} p+T+\tilde{n}+1)/2}\exp\bigg\{-\frac{1}{2}\text{tr}\Big\{\Big(V_0\\
&+&\big(\Pi-\Pi_0\big)\Lambda_0^{-1}\big(\Pi-\Pi_0\big)'+\big(\bar{y}-\Pi\bar{\mathsf{Y}}\big)\big(\bar{y}-\Pi\bar{\mathsf{Y}}\big)'\Big)\Sigma^{-1}\Big\}\bigg\}.\nonumber
\end{eqnarray}
Let us consider the sum of the terms corresponding to the prior density of the parameter $\Pi$ and the likelihood function in the last line of the above equation. Then, it can be shown that
\begin{eqnarray}\label{02050}
&&(\Pi-\Pi_0)\Lambda_0^{-1}(\Pi-\Pi_0)'+(\bar{y}-\Pi\bar{\mathsf{Y}})(\bar{y}-\Pi\bar{\mathsf{Y}})'\nonumber\\
&&=(\Pi-\Pi_{*|T})(\Lambda_0^{-1}+\bar{\mathsf{Y}}\bar{\mathsf{Y}}')(\Pi-\Pi_{*|T})'\\
&&-\Pi_{*|T}(\Lambda_0^{-1}+\bar{\mathsf{Y}}\bar{\mathsf{Y}}')\Pi_{*|T}'+\Pi_0\Lambda_0^{-1}\Pi_0'+\bar{y}\bar{y}',\nonumber
\end{eqnarray}
where $\Pi_{*|T}=(\Pi_0\Lambda_0^{-1}+\bar{y}\bar{\mathsf{Y}}')(\Lambda_0^{-1}+\bar{\mathsf{Y}}\bar{\mathsf{Y}}')^{-1}$. Consequently, according to equation \eqref{02050}, the posterior density of the parameter $(\Pi,\Sigma)$ takes form of a multivariate normal density times an inverse--Wishart density
\begin{equation}\label{02051}
f(\Pi,\Sigma|\bar{y},\bar{\mathsf{Y}})=f\big(\pi\big|\Sigma,\pi_{*|T},\Lambda_{*|T},\bar{y},\bar{\mathsf{Y}}\big)f\big(\Sigma\big|\nu_*,V_{*|T},\bar{y},\bar{\mathsf{Y}}\big)
\end{equation}
where
\begin{eqnarray}\label{02052}
&&f\big(\pi\big|\Sigma,\pi_{*|T},\Lambda_{*|T},\bar{y},\bar{\mathsf{Y}}\big)\\
&&=\frac{1}{(2\pi)^{[\tilde{n}(l+\tilde{n}p)]/2}|\Lambda_{*|T}|^{\tilde{n}/2}|\Sigma|^{(l+\tilde{n}p)/2}}\exp\bigg\{-\frac{1}{2}\Big(\pi-\pi_{*|T}\Big)'\big(\Sigma^{-1}\otimes \Lambda_{*|T}^{-1}\big)\Big(\pi-\pi_{*|T}\Big)\bigg\}\nonumber
\end{eqnarray}
with
\begin{equation}\label{02053}
\pi_{*|T}:=\text{vec}(\Pi_{*|T})~~~\text{and}~~~\Lambda_{*|T}^{-1}:=\Lambda_0^{-1}+\bar{\mathsf{Y}}\bar{\mathsf{Y}}'
\end{equation}
and
\begin{equation}\label{02054}
f\big(\Sigma\big|\nu_*,V_{*|T},\bar{y},\bar{\mathsf{Y}})=\frac{|V_{*|T}|^{\nu_*/2}}{2^{\tilde{n}\nu_*/2}\Gamma_{\tilde{n}}(\nu_*/2)}|\Sigma|^{-(\nu_*+\tilde{n}+1)/2}\exp\bigg\{-\frac{1}{2}\text{tr}\Big(V_{*|T}\Sigma^{-1}\Big)\bigg\}
\end{equation}
with 
\begin{eqnarray}
\nu_*&:=&\nu_0+l+\tilde{n}p+T \label{02055}\\
V_{*|T}&:=&V_0-\Pi_{*|T}(\Lambda_0^{-1}+\bar{\mathsf{Y}}\bar{\mathsf{Y}}')\Pi_{*|T}'+\Pi_0\Lambda_0^{-1}\Pi_0'+\bar{y}\bar{y}'\label{02056}
\end{eqnarray}
Note that if $\Lambda_0^{-1}\to 0$, which corresponds to uninformative diffuse prior, then the posterior mean \eqref{02053} converges to the maximum likelihood estimator $\hat{\Pi}= \bar{y}\bar{\mathsf{Y}}'(\bar{\mathsf{Y}}\bar{\mathsf{Y}}')^{-1}$. By the tower property of conditional expectation, the Bayesian estimator of the parameter vector $\Pi$ is obtained by
\begin{equation}\label{02057}
\Pi_{*|T}=\mathbb{E}(\Pi|\bar{y},\bar{\mathsf{Y}})=(\Pi_0\Lambda_0^{-1}+\bar{y}\bar{\mathsf{Y}}')(\Lambda_0^{-1}+\bar{\mathsf{Y}}\bar{\mathsf{Y}}')^{-1}.
\end{equation}
Due to the expectation formula of inverse--Wishart distribution, the Bayesian estimator of the parameter $\Sigma$ is given by
\begin{equation}\label{02058}
\Sigma_{*|T}:=\mathbb{E}(\Sigma|\bar{y},\bar{\mathsf{Y}})=\frac{1}{\nu_*-\tilde{n}-1}V_{*|T}.
\end{equation}

To make statistical inferences about the parameter vector $\theta=(\Pi,\Sigma)$ conditional on the information $\bar{y}$ and $\bar{\mathsf{Y}}$, one may use the Gibbs sampling method, which generates a dependent sequence of our parameters. In the Bayesian statistics, the Gibbs sampling is often used when the joint distribution is not known explicitly or is difficult to sample from directly, but the conditional distribution of each variable is known and is easy to sample from. Constructing the Gibbs sampler to approximate the joint posterior distribution $f(\Pi,\Sigma|\bar{y}_T,\bar{\mathsf{Y}}_T)$ given in equation \eqref{02051} is straightforward: New
values $\big(\pi_{(s)}, \Sigma_{(s)}\big)$, $s=1,\dots,N$ can be generated by
\begin{itemize}
\item[1.] sample $\Sigma_{(s)}\sim \mathcal{IW}(\nu_*,V_{*|T})$ 
\item[2.] sample $\pi_{(s)}\sim \mathcal{N}\big(\pi_{*|T},\Sigma_{(s)}\otimes\Lambda_{*|T}\big),$
\end{itemize}
where $\mathcal{IW}$ is an abbreviation of the inverse--Wishart distribution, and the parameters $\nu_*$ and $V_{*|T}$ of the inverse--Wishart distribution and mean $\pi_{*|T}$ and the matrix $\Lambda_{*|T}$ of the multivariate normal distribution are given in equations \eqref{02055}--\eqref{02056} and \eqref{02053}, respectively. 

As mentioned before, VARs tend to have a lot of parameters, and large VARs exacerbate this problem. In particular, for a VAR($p$) process with order of $p=3$, $l=1$ exogenous variable and $\tilde{n}=15$ endogenous variables, we have to estimate $\tilde{n}(l+\tilde{n}p)=690$ VAR coefficients. In this case, the number of VAR coefficients is much larger than the number of observations for small and medium--sized samples. Therefore, without informative priors or regularization, it is not even possible to estimate the VAR coefficients. 

In practice, one usually adopts Minnesota prior to estimating the parameters of the VAR$(p)$ process. \citeA{Doan84} firstly introduced Minnesota prior to small Bayesian VAR. Also, \citeA{Banbura10} used Minnesota prior for large Bayesian VAR and showed that the forecast of large Bayesian VAR is better than small Bayesian VAR. However, there are many different variants of the Minnesota prior, to illustrative purposes, we consider the prior, which is included in \citeA{Banbura10}. 

The idea of Minnesota prior is that it shrinks diagonal elements of the matrix $A_1$ toward $\delta_i$ and off--diagonal elements of $A_1$ and all elements of other matrices $A_0,A_2,\dots,A_p$ toward 0, where $\delta_i$ is 0 for a stationary variable $y_{i,t}$ and 1 for a variable with unit root $y_{i,t}$.  For the prior, it is assumed that conditional on $\Sigma$, $A_0,A_1,\dots,A_p$ are independent, normally distributed, and for $(i,j)$--th element of the matrix $A_s$ $(s=0,\dots,p)$ it holds
\begin{equation}\label{02059}
\mathbb{E}\big((A_s)_{ij}\big|\Sigma\big)=\begin{cases}
\delta_i & \text{if}~~~i=j,~s=1,\\
0 & \text{if}~~~\text{otherwise}
\end{cases},
\end{equation}
\begin{equation}\label{02060}
\text{Var}\big((A_0)_{ij}\big|\Sigma\big)=1/\varepsilon_{ij}, ~~\text{and}~~
\text{Var}\big((A_s)_{ij}\big|\Sigma\big)=\begin{cases}
\frac{\lambda^2}{s^2} & \text{if}~~~i=j,\\
\theta\frac{\lambda^2}{s^2}\frac{\sigma_i}{\sigma_j} & \text{if}~~~\text{otherwise}
\end{cases}~~\text{for}~s=1,\dots,p,
\end{equation}
where we denote $(i,j)$--th element of the matrix $A_s$ by $(A_s)_{ij}$. 

The parameter $\varepsilon_{ij}$ is a small number and it corresponds to an uninformative diffuse prior for $(A_0)_{ij}$, the parameter $\lambda$ controls the overall tightness of the prior distribution, the factor $1/s^2$ is the rate at which prior variance decreases with increasing lag length, the factor $\sigma_i/\sigma_j$ accounts for different scale and variability of the data, and the coefficient $\theta\in[0,1]$ governs the extent to which the lags of the other variables are less important than the own lags. By using the dummy variables, \citeA{Banbura10} obtain Bayesian estimators, corresponding to the hyperparameters.  

For our Bayesian estimators, given in equation \eqref{02057} and \eqref{02058}, we can not use Minnesota prior due to the Kronecker product, $\Sigma\otimes \Lambda_0$. For this reason, to define a prior, which applies the idea of Minnesota prior, we follow \citeA{Chan20}. One should note that $\pi_0$, $\Lambda_0$, $\nu_0$, and $V_0$ are hyper--parameters of our model. For the hyperparameter $\nu_0$, to the prior variance of $\Sigma$ is large, which corresponds relatively uninformative diffuse prior, it is often chosen small value for $\nu_0$. According to the expectation formula of the inverse--Wishart distribution, we have $\mathbb{E}(\Sigma)=\frac{1}{\nu_0-\tilde{n}-1}V_0$. Consequently, for given $\nu_0$, one chooses $V_0$ to match the desired prior mean of $\Sigma$ using the expectation formula. For the hyperparameter $\pi_0$, one may use equation \eqref{02059}. To introduce shrinkage in the hyperparameter $\Lambda_0$, we assume that it is a diagonal matrix. Then, its diagonal elements are
\begin{equation}\label{02061}
\Lambda_{0,ii}=\begin{cases}
\lambda_1 & \text{if}~~~1\leq i\leq l,\\
\displaystyle \frac{\lambda_2}{s^2\sigma_s^2} & \text{if}~~~l+\tilde{n}(s-1)<i\leq l+\tilde{n}s,~s=1,\dots,p
\end{cases}
\end{equation}
for $i=1,\dots l+\tilde{n}p$. Some other priors and Bayesian methods can be found in \citeA{Chan20}.

In practice, to estimate the parameters $\sigma_1^2,\dots,\sigma_{\tilde{n}}^2$, for $i=1,\dots,\tilde{n}$, we model each individual variable $y_{i,t}$ by univariate autoregressive model with order $p$ (AR$(p)$). Then, we estimate the AR$(p)$ processes by the ordinary least square (OLS) method. If we denote standard OLS estimate of the error variance of the $i$--th AR($p$) equation by $s_i^2$, then the parameter $\sigma_i^2$ is estimated by $s_i^2$. 

\subsection{Portfolio Selection for Public Company}\label{sub03}

The mean--variance portfolio choice model was established firstly by \citeA{Markowitz52}. In the stochastic DDM framework, by introducing a discrete joint distribution for dividend growth rates, the first time \citeA{Agosto19} obtained a closed--form covariance formula between two stocks. Also, they consider the portfolio choice problem for two stocks. Furthermore, using a multivariate Markov chain, \citeA{dAmico20b} provided a portfolio selection of three stocks. \citeA{Battulga22a} obtained higher order moments of stock price by applying non homogeneous Poisson process.

In this Subsection, we consider a problem that a public company has some cash at time 0 and wants to maximize its mean--variance utility function on the next period's earnings before tax, which comes from buying stocks, including its own, and paying interest payments on liabilities. Then, the problem is given by the following portfolio choice problem with the mean--variance utility function 
\begin{equation}\label{02062}
\begin{cases}
\displaystyle\mathbb{E}\big(x'k_1^e-x_ik_{i,0}^\ell\big|\mathcal{F}_0\big)-\frac{c}{2}\text{Var}\big(x'k_1^e-x_ik_{i,0}^\ell\big|\mathcal{F}_0\big)\longrightarrow\max\\
\text{s.t.}~x'i_n+x_i=1,
\end{cases}
\end{equation}
where $k_1^e=(k_{1,1}^e,\dots,k_{n,1}^e)'$ is an $(n\times 1)$ vector, consisting of the required rate of returns at time 1 on equities, $k_{i,0}^\ell$ is the required rate of return at time 0 on liabilities of $i$--th company, calculated by equation \eqref{02009}, $(x',x_i)'$ is an $([n+1]\times 1)$ variables' vector, and $c>0$ is a risk--aversion parameter, which is different for each investor. The problem is equivalent to the following problem
\begin{equation}\label{02063}
\begin{cases}
\displaystyle x'\mu-x_ik_{i,0}^\ell-\frac{c}{2}x'\Sigma x\longrightarrow\max\\
\text{s.t.}~x'i_n+x_i=1,
\end{cases}
\end{equation}
where $\mu:=\mathbb{E}(k_1^e|\mathcal{F}_0)=\mathbb{E}\big(\exp\{Jy_1\}\big|\mathcal{F}_0\big)$ and $\Sigma:=\text{Var}(k_1^e|\mathcal{F}_0)=\text{Var}\big(\exp\{Jy_1\}\big|\mathcal{F}_0\big)$  are $(n\times 1)$ conditional expectation vector and $(n\times n)$ conditional covariance matrix of the required rate of return vector on equities $k_1^e$, respectively, and $J:=[I_n:0]$ is an $(n\times \tilde{n})$ matrix, which is used to extract the log required rate of return vector $\tilde{k}_1^e$ from the vector $y_1$. The problem is the quadratic programming problem and it has a unique solution. Its Lagrangian function is given by
\begin{equation}\label{02064}
\mathcal{L}(x,x_i,\lambda):=x'\mu-x_ik_{i,0}^\ell-\frac{c}{2}x'\Sigma x-\lambda(x'i_n+x_i-1).
\end{equation}
Taking partial derivatives with respect to parameters $x$, $x_i$, and $\lambda$ from the Lagrangian function and setting to zero, one finds the solution of the quadratic programming problem
\begin{equation}\label{02065}
x^*:=\frac{1}{c}\Sigma^{-1}\big(\mu+k_{i,0}^\ell i_n\big)~~~\text{and}~~~x_i^*:=1-\frac{1}{c}i_n'\Sigma^{-1}\big(\mu+k_{i,0}^\ell i_n\big).
\end{equation}
To obtain a solution to the problem \eqref{02063}, we need to calculate the conditional expectation vector $\mu_i$ and conditional covariance matrix $\Sigma_i$. 

According to the expectation and covariance formula of the log--normal random vector and the fact that $\mathbb{E}(y_1|\mathcal{F}_0,s_1)=\Pi(s_1)\mathsf{Y}_0$ and $\text{Var}(y_1|\mathcal{F}_0,s_1)=\Sigma_1(s_1)$, we have that
\begin{equation}\label{02066}
\mathbb{E}\big(\exp\{Jy_1\}\big|\mathcal{F}_0,s_1\big)=\exp\bigg\{J\Pi(s_1)\mathsf{Y}_0+\frac{1}{2}J\Sigma_1(s_1)J'\bigg\}
\end{equation}  
and
\begin{eqnarray}\label{02067}
&&\text{Var}\big(\exp\{Jy_1\}\big|\mathcal{F}_0,s_1\big)=\Bigg(\exp\bigg\{J\Pi(s_1)\mathsf{Y}_0+\frac{1}{2}J\Sigma_1(s_1)J'\bigg\}\\
&&\times \exp\bigg\{J\Pi(s_1)\mathsf{Y}_0+\frac{1}{2}J\Sigma_1(s_1)J'\bigg\}'\Bigg)\odot\bigg(\exp\Big\{J\Sigma_1(s_1)J'\Big\}-\mathsf{E}_n\bigg),\nonumber
\end{eqnarray}  
where $\mathsf{E}_n$ is an $(n\times n)$ matrix, whose elements equal 1. As a result, we get the parameters $\mu_i$ and $\Sigma_i$ in equation \eqref{02065}, corresponding to the portfolio selection with regime-switching:
\begin{equation}\label{02068}
\mu=\sum_{s_1=1}^N\mathbb{E}\big(\exp\{Jy_1\}\big|\mathcal{F}_0,s_1\big)p_{s_1}~~~\text{and}~~~\Sigma=\sum_{s_1=1}^N\text{Var}\big(\exp\{J_iy_1\}\big|\mathcal{F}_0,s_1\big)p_{s_1}.
\end{equation}

The solution of the problem is not only maximizing the earning before tax but also optimizing a capital structure of a company. Let us assume that $i$--th company has some cash, say \$50 million. Then, the company may optimize its capital structure of the balance sheet, namely, ($i$) reduce liabilities or expand the liabilities by $50\times x_i^*$ and ($ii$) reduce treasury stocks of the company or expand the stocks by $50\times \bar{x}_i^*$, where $\bar{x}_i^*$ is $i$--th component of the optimal vector $x^*$. Of course, one may add constraints to the problem to prohibit short sales.

\section{Parameter Estimation of Private Company}

In this Section, we will consider parameter estimation methods for $n$ private companies. Let $B_{t}$ be a book value of equity and $b_{t}$ be a book value growth rate, respectively, at time $t$ of a private company. Since the book value of equity at time $t-1$ grows at rate $b_{t}$, its value at time $t$ becomes 
\begin{equation}\label{02072}
B_{t}=(1+b_{t})B_{t-1}.
\end{equation}
If we assume that for the private company, its price--to--book ratio is constant, say, $m=P_{t}/B_{t}$, for all $t=1,\dots,T$, then according to DDM equation \eqref{02001}, price (value) at time $t$ of the private company is expressed by the following equation
\begin{equation}\label{02073}
mB_{t}=(1+k_{t})mB_{t-1}-d_{t}=\big((1+k_{t})m-\Delta_{t}\big)B_{t-1},
\end{equation}
where $k_{t}$ is the required rate of return on equity at time $t$ and $\Delta_{t}:=d_{t}/B_{t-1}$ is a dividend--to--book ratio at time $t$, respectively, of the private company. If we substitute equation \eqref{02072} into the left--hand side of the above equation \eqref{02073}, then we get that
\begin{equation}\label{02074}
(1+b_{t})mB_{t-1}=\big((1+k_{t})m-\Delta_{t}\big)B_{t-1}.
\end{equation}
Therefore, a relation between the dividend--to--book ratio, book value growth rate, required rate of return, and the price--to--book ratio is given by
\begin{equation}\label{02075}
mb_{t}=m k_{t}-\Delta_{t}, ~~~t=1,2,\dots.
\end{equation}
We refer to the model and its versions with the regime--switching and with a state (latent or unobserved) variable given in equations \eqref{02077} and \eqref{02093}, respectively, as the private company valuation model. For the log private company valuation model, we refer to \citeA{Battulga23c}, where he considers the private company valuation model in the log framework, and obtain closed--form pricing and hedging formulas for European call and put options. It should be noted that the private company valuation model given in \eqref{02075} is equivalent to the franchise factor model, see \citeA{Leibowitz90}, but the private company valuation models with the regime--switching and the state variable differs from the franchise factor model. According to equation \eqref{02075}, the required rate of return at time $t$ of the private company is represented by
\begin{equation}\label{02076}
k_{t}=\frac{1}{m}\Delta_{t}+b_{t}.
\end{equation}
From the above equation, one can see that for a dividend--paying private company, if $m$ increases, then the required rate of return $k_{t}$ decreases and it converges to the book value growth rate $b_{t}$. Thus, as the price--to--book and dividend--to--book ratios are positive, the book value growth rate is a floor of the required rate of return. On the other hand, because the term $\frac{1}{m}\Delta_{t}$ takes positive value, if a private company pays dividends, then its required rate of return is always greater than the does not pay case.

\subsection{Regime Switching Estimation}\label{sub04}

In order to incorporate a case, where the dividends are not paid into maximum likelihood estimators of the private company valuation model's parameter, rather than equation \eqref{02075} we will use equation \eqref{02076}. As some private companies may not pay dividends, we suppose that there are $n_d$ ($0\leq n_d\leq n$) companies that pay dividends. Because it is always possible to change the order of the companies, without loss of generality we can assume that dividend--paying companies are placed first $n_d$ components of the first equation of system \eqref{02080}, see below. 

To keep notations simple, let $B_t:=(B_{1,t},\dots,B_{n,t})'$ be an $(n\times 1)$ book value vector, $m(s_t):=(m_1(s_t),\dots,m_{n_d}(s_t))'$ be an $(n_d\times 1)$ price--to--book ratio vector in regime $s_t$ corresponding to dividend paying companies, $b_t:=(b_{1,t},\dots,b_{n,t})'$ be an $(n\times 1)$ book value growth rate vector, and $k_t(s_t):=(k_{1,t}(s_t),\dots,k_{n,t}(s_t))'$ be an $(n\times 1)$ required rate of return vector in regime $s_t$, $r(s_t):=(1/m_1(s_t),\dots,1/m_{n_d}(s_t))'$ be an $(n\times 1)$ book--to--price ratio vector in regime $s_t$ and reciprocal of the vector $m(s_t)$, and $R(s_t):=\text{diag}\{r(s_t),0\}$ be an $(n\times n)$ diagonal matrix, whose diagonal elements consist of the book--to--price ratio vector $r(s_t)$ and an $((n-n_d)\times 1)$ zero vector. Then, equation \eqref{02076} can be written by
\begin{equation}\label{02077}
b_t=k_t(s_t)-R(s_t)\Delta_{t}.
\end{equation}

Since the book value growth rate process $b_t$ may depend on economic variables, we define an $(\ell\times 1)$ MS--VAR($p$) process $x_t$
\begin{equation}\label{02078}
x_t=A_0^x(s_t)\psi_t+A_1^x(s_t)x_{t-1}+\dots+A_p^x(s_t)x_{t-p}+v_t,
\end{equation}
where $x_t=(x_{1,t},\dots,x_{\ell,t})'$ is an $(\ell\times 1)$ random vector, $\psi_t=(\psi_{1,t},\dots,\psi_{l,t})'$ is a $(l\times 1)$ random vector of exogenous variables, $v_t=(u_{1,t},\dots,v_{\ell,t})'$ is an $(\ell\times 1)$ residual process, $s_t$ is an unobserved regime at time $t$, which is governed by a Markov chain with $N$ states, $A_0^x(s_t)$ is an $(\ell\times l)$ coefficient matrix at regime $s_t$ that corresponds to the vector of exogenous variables, for $i=1,\dots,p$, $A_i^x(s_t)$ are $(\ell\times \ell)$ coefficient matrices at regime $s_t$ that correspond to $x_{t-1},\dots,x_{t-p}$. The process $x_t$ consists of the economic variables that affect the book value growth rate process $b_t$. Note that the process $x_t$ can include dividend--to--book ratio process $\Delta_t$. Equation \eqref{02078} can be compactly written by
\begin{equation}\label{02079}
x_t=\Pi^x(s_t)\mathsf{X}_{t-1}+v_t,
\end{equation}
where $\mathsf{X}_{t-1}:=(\psi_t',x_{t-1}',\dots,x_{t-p}')'$ is an $\big([l+\ell p]\times 1\big)$ vector, which consists of exogenous variables and last $p$ lagged values of the process $x_t$ and $\Pi^x(s_t):=[A_0^x(s_t):A_1^x(s_t):\dots:A_p^x(s_t)]$ is an $\big(\ell\times [l+\ell p]\big)$ matrix, which consists of the coefficient matrices of the process $x_t$. We suppose that the required rate of return depends on the exogenous variables and random amount $u_t$, namely, $k_t(s_t)=A_0^k(s_t)\psi_t+u_t$. Consequently, our private company valuation model is given by the following system
\begin{equation}\label{02080}
\begin{cases}
b_t=A_0^k(s_t)\psi_t-R(s_t)\Delta_{t}+u_t\\
x_t=\Pi^x(s_t)\mathsf{X}_{t-1}+v_t
\end{cases}.
\end{equation}

To simplify the model, we assume that a covariance matrix of a random residual process $\xi_t:=(u_t',v_t')'$ is homoscedastic, that is, $\text{Var}(\xi_t)=\Sigma(s_t)$. However, one can easily develop private company valuation models with heteroscedastic residuals as in Section 2. If regime random vector $s_t$ is in a regime $j$, then conditional on the information $\mathcal{F}_{t-1}$, a log conditional density of a random vector $y_t:=(b_t',x_t')'$ is given by
\begin{eqnarray}\label{02081}
\ln(\eta_{tj})&=&\ln\big(f(y_t|s_t=j,\mathcal{F}_{t-1},\alpha)\big)=-\frac{\tilde{n}}{2}\ln(2\pi)-\frac{1}{2}\ln(|\Sigma(j)|)\\
&-&\frac{1}{2}\Big(u_t'(j)\Omega_{uu}(j)u_t(j)+2u_t'(j)\Omega_{uv}(j)v_t(j)+v_t'(j)\Omega_{vv}(j)v_t(j)\Big),\nonumber
\end{eqnarray}
where the residual vectors in the regime $j$ are $u_t(j)=b_t-A_0^k(j)\psi_t+R(j)\Delta_{t}$ and $v_t(j)=x_t-\Pi^x(j)\mathsf{X}_{t-1}$ and $\Omega_{uu}(j)$, $\Omega_{uv}(j)$, $\Omega_{vu}(j)$, and $\Omega_{vv}(j)$ are partitions, corresponding to a residual vector $\xi_t(j):=(u_t'(j),v_t'(j))'$ of a matrix $\Sigma(j)^{-1}$. 

To obtain the partial derivative of the log conditional density with respect to the book--to--price ratio vector $r(s_t)$, instead of the first equation of system \eqref{02080}, we need an equation $J_db_t=J_dA_0^k(s_t)\psi_t+J_d\text{diag}{\{\Delta(s_t)\}}J_d'r(s_t)+J_du_t$, where $J_d:=[I_{n_d}:0]$ is an $(n_d\times n)$ matrix. Consequently, the partial derivative is given by
\begin{equation}\label{02082}
\frac{\partial \ln(\eta_{tj})}{\partial r(j)'}=-\Big(u_t'(j)J_d'J_d\Omega_{uu}J_d'+v_t'(j)\Omega_{vu}J_d'\Big)J_d\text{diag}\{\Delta_t\}J_d'.
\end{equation} 
By equation \eqref{02024}, for each regime $j=1,\dots,N$, one obtains ML estimator of the parameter vector $r(j)$
\begin{eqnarray}\label{02083}
\hat{r}(j)&:=&\Bigg(\sum_{t=1}^TJ_d\text{diag}\{\bar{\Delta}_{t,j}\}J_d'J_d\Omega_{uu}(j)J_dJ_d'\text{diag}\{\bar{\Delta}_{t,j}\}J_d'\Bigg)^{-1}\nonumber\\
&\times&\sum_{t=1}^TJ_d\text{diag}\{\bar{\Delta}_{t,j}\}J_d'\bigg(J_d\Omega_{uu}(j)J_d'J_d\Big(A_0^k(j)\bar{\psi}_{t,j}-\bar{b}_{t,j}\Big)\\
&-&J_d\Omega_{uv}(j)\Big(\bar{x}_{t,j}-\Pi^x(j)\bar{\mathsf{X}}_{t-1,j}\Big)\bigg),\nonumber
\end{eqnarray}
where $\bar{\Delta}_{t,j}:=\Delta_t\sqrt{(z_{t|T})_j}$ is an $(n\times 1)$ dividend--to--book ratio process, adjusted by the regime $j$, $\bar{\psi}_{t,j}:=\psi_t\sqrt{(z_{t|T})_j}$ is an $(l\times 1)$ exogenous variables vector, adjusted by the regime $j$, $\bar{b}_{t,j}:=b_t\sqrt{(z_{t|T})_j}$ is an $(n\times 1)$ book value growth rate process, adjusted by the regime $j$, and $\bar{\mathsf{X}}_{t,j}:=\mathsf{X}_t\sqrt{(z_{t|T})_j}$ is an $\big([l+\ell p]\times 1\big)$ explanatory variables vector, adjusted by the regime $j$.

Let $a^k(j):=\text{vec}\big(A_0^k(j)\big)$ be a vectorization of the matrix $A_0^k(j)$. Then, as $A_0^k(j)\psi_t=(\psi_t'\otimes I_n)a^k(j)$ and partial derivative of the log conditional density with respect to the vector $a^k(j)$ is
\begin{equation}\label{02084}
\frac{\partial \ln(\eta_{tj})}{\partial a^k(j)'}=\Big(u_t'(j)\Omega_{uu}+v_t'(j)\Omega_{vu}\Big)(\psi_t'\otimes I_n).
\end{equation} 
According to equation \eqref{02084} and the procedure, which is used to obtain equations \eqref{02033} and \eqref{02035}, for each regime $j=1,\dots,N$, we obtain ML estimator of the parameter matrix $A_0^k(j)$
\begin{equation}\label{02085}
\hat{A}_0^k(j):=\Big(\bar{b}_j+R(j)\bar{\Delta}_j+\Omega_{uu}^{-1}(j)\Omega_{uv}(j)\big(\bar{x}_j-\Pi(j)\mathsf{X}_j\big)\Big)\bar{\psi}_j'\big(\bar{\psi}_j\bar{\psi}_j'\big)^{-1},
\end{equation}
where $\bar{b}_j:=[\bar{b}_{1,j}:\dots:\bar{b}_{T,j}]$ is an $(n\times T)$ matrix, $\bar{\Delta}_j:=[\bar{\Delta}_{1,j}:\dots:\bar{\Delta}_{T,j}]$ is an $(n\times T)$ matrix, $\bar{\mathsf{X}}_j:=[\bar{\mathsf{X}}_{0,j}:\dots:\bar{\mathsf{X}}_{T-1,j}]$ is an $([l+\ell p]\times T)$ matrix, and $\bar{\psi}_j:=[\bar{\psi}_{1,j}:\dots:\bar{\psi}_{T,j}]$ is an $(l\times T)$ matrix.

Similarly, for each regime $j=1,\dots,N$, one finds ML estimator of the parameter matrix $\Pi^x(j)$:
\begin{equation}\label{02086}
\hat{\Pi}^x(j):=\Big(\bar{x}_j+\Omega_{vv}^{-1}(j)\Omega_{vu}(j)\big(\bar{b}_j-A_k(j)\bar{\psi}_j+R(j)\bar{\Delta}_j\big)\Big)\bar{\mathsf{X}}_j'\big(\bar{\mathsf{X}}_j\bar{\mathsf{X}}_j'\big)^{-1}.
\end{equation}

Analogous to equation \eqref{02035}, it can be shown that for each regime $j=1,\dots,N$, ML estimator of the covariance matrix $\Sigma(j)$ is given by
\begin{equation}\label{02087}
\hat{\Sigma}(j)=\frac{1}{\sum_{t=1}^T(z_{t|T})_j}\begin{bmatrix}
\bar{u}_j\bar{u}_j' & \bar{u}_j\bar{v}_j'\\
\bar{v}_j\bar{u}_j' & \bar{v}_j\bar{v}_j'
\end{bmatrix}
\end{equation}
where the residual matrices that are adjusted by the regime $j$ are $\bar{u}_j:=\bar{b}_j-A_0^k(j)\bar{\psi}_j+R(j)\bar{\Delta}_j$ and $\bar{v}_j:=\bar{x}_j-\Pi^x(j)\bar{\mathsf{X}}_j$. 

It is worth mentioning that if all the companies do not pay dividends, then for each $j=1,\dots,N$, we do not need to estimate the parameter $r(j)$. Consequently, ML estimators of the parameters $A_0^k(j)$ and $\Pi^x(j)$ are obtained by substituting $\bar{\Delta}_j=0$ into equations \eqref{02085} and \eqref{02086}. 
  
\subsection{The Bayesian Estimation}\label{sub05}

Now, we move to the Bayesian analysis of linear regression. To obtain the Bayesian estimator of the private company valuation model, we need the following multivariate linear regression that corresponds to system \eqref{02080}
\begin{equation}\label{02088}
y_t=\Pi\mathsf{Y}_{t-1}+\xi_t,
\end{equation}
where $y_t:=(b_t',x_t')'$ is an $(\tilde{n}\times 1)$ vector, $C$ is an $(\tilde{n}\times [n+l+\ell p])$ random matrix, $\mathsf{Y}_{t-1}:=\big(\Delta_t',\psi_t',x_{t-1}',\dots,x_{t-p}'\big)'$ is an $([n+l+\ell p]\times 1)$ vector, and $\xi_t:=(u_t',v_t')'$ is an $(\tilde{n}\times 1)$ white noise process with a random covariance matrix $\Sigma=\text{Var}(\xi_t)$. The matrix $C$ has the following structure
\begin{equation}\label{02089}
\Pi=\begin{bmatrix}
\Pi_{b_t\Delta_t} & \Pi_{b_t\psi_t} & \Pi_{b_tx_{t-1}} & \Pi_{b_tx_{t-2}} & \dots & \Pi_{b_tx_{t-p}}\\
\Pi_{x_t\Delta_t} & \Pi_{x_t\psi_t} & \Pi_{x_tx_{t-1}} & \Pi_{x_tx_{t-2}} & \dots & \Pi_{x_tx_{t-p}}
\end{bmatrix},
\end{equation}
where for $\alpha\in\{b_t,x_t\}$ and $\beta\in\{\Delta_t,\psi_t,x_{t-1},\dots,x_{t-p}\}$, $\Pi_{\alpha\beta}$ is a random coefficient matrix of the random vector $\beta$, corresponding to the process $\alpha$.
Taking into account the structure of system \eqref{02080}, we expect a prior expectation matrix of the random matrix $\Pi$ is given by
\begin{eqnarray}\label{02090}
\Pi_0=\mathbb{E}\big[\Pi\big|\Sigma\big]&=&\begin{bmatrix}
\Pi_{b_t\Delta_t}^* & \Pi_{b_t\psi_t}^* & 0 & 0 & \dots & 0\\
0 & 0 & \Pi_{x_tx_{t-1}}^* & 0 & \dots & 0
\end{bmatrix},
\end{eqnarray}
where $\Pi_{b_t\Delta_t}^*:=\mathbb{E}\big[\Pi_{b_t\Delta_t}\big|\Sigma\big]$ is an $(n\times n)$ diagonal matrix, whose first $n_d$ components are correspond to prior expectation of the book--to--price ratio vector $r$ and other components are zero, $\Pi_{b_t\psi_t}^*:=\mathbb{E}\big[\Pi_{b_t\psi_t}\big|\Sigma\big]$ is an $(n\times l)$ prior expectation matrix of the random matrix $A_0^k$, and $\Pi_{x_tx_{t-1}}^*:=\mathbb{E}\big[\Pi_{x_tx_{t-1}}\big|\Sigma\big]$ is an $(\ell\times \ell)$ diagonal prior expectation matrix of the random matrix $A_1^x$ and its diagonal elements are given by equation \eqref{02059}. To obtain the prior variance of the random matrix $\Pi$, we apply the idea in equation \eqref{02061}. By using the idea, diagonal elements of $([n+l+\ell p]\times [n+l+\ell p])$ diagonal matrix $\Lambda_0$ are defined by
\begin{equation}\label{02091}
\Lambda_{0,ii}=\begin{cases}
\lambda_1 & \text{if}~~~1\leq i\leq n+l,\\
\displaystyle\frac{\lambda_2}{s^2\sigma_s^2} & \text{if}~~~n+l+\ell(s-1)<i\leq n+l+\ell s,~s=1,\dots,p. 
\end{cases}
\end{equation}
for $i=1,\dots,n+l+\ell p$. Other hyperparameters $\nu_0$ and $V_0$ are exactly the same defined as in Section \ref{sub02}. After defining the hyperparameters, one can obtain the Bayesian estimators using equations \eqref{02057} and \eqref{02058}.

\subsection{The Kalman Filtering}\label{sub06}

Because one can use ideas, which arise in the following to estimate the required rate of return on debtholders, in this Subsection, we will concentrate on the required rate of return on equity. Let us assume that the price--to--book ratio varies over time, that is, $m_t=P_t/B_t$, $t=1,\dots,T$. Under the assumption, for a generic private company, equation \eqref{02074} becomes
\begin{equation}\label{02092}
m_tB_t=\big((1+k_t^\circ)m_{t-1}-\Delta_t\big)B_{t-1}.
\end{equation} 
Therefore, using the relation $B_t=(1+b_t)B_{t-1}$ in equation \eqref{02092} a relation between the dividend--to--book ratio, book value growth rate, the required rate of return on equity, and price--to--book ratios is given by
\begin{equation}\label{02093}
\Delta_t=-(1+b_t)m_t+(1+k_t^\circ)m_{t-1}.
\end{equation}
To estimate the parameters of the required rate of return on equity, we must add a random amount, say, $u_t$, into equation \eqref{02093}. Then, equation \eqref{02093} becomes
\begin{equation}\label{02094}
\Delta_t=-(1+b_t)m_t+(1+k_t^\circ)m_{t-1}+u_t.
\end{equation}
It should be noted that for the above equation, the price--to--book ratios $m_t$ and $m_{t-1}$ are unobserved (state) variables. For a non--dividend paying firm, the above equation becomes
\begin{equation}\label{02095}
\tilde{b}_t=\tilde{k}_t^\circ-\tilde{m}_t+\tilde{m}_{t-1}+u_t,
\end{equation}
where $\tilde{b}_t:=\ln(1+b_t)$ is a log book value growth rate, $\tilde{k}_t^\circ:=\ln(1+k_t^\circ)$ is a log required rate of return on equity, and $\tilde{m}_t:=\ln(m_t)$ is an unobserved log price--to--book ratio, respectively, at time $t$ of the non--dividend paying company. 

We assume that the price--to--book ratio and log price--to--book ratio are governed by the autoregressive distributed lag model of order $(q,p)$ (ADL$(q,p)$), that is, 
\begin{eqnarray}\label{02096}
m_t&=&\Phi_1m_{t-1}+\dots+\Phi_qm_{t-q}+A_0^m\psi_t+A_1^mx_{t-1}+\dots+A_p^mx_{t-p}+w_t\\
&=&\Phi\mathsf{M}_{t-1}+\Pi^x\mathsf{X}_{t-1}+w_t\nonumber
\end{eqnarray}
and 
\begin{equation}\label{02097}
\tilde{m}_t=\Phi\tilde{\mathsf{M}}_{t-1}+\Pi^x\mathsf{X}_{t-1}+w_t
\end{equation}
where for $i=1,\dots,n$, $\Phi_i$ is an $(n\times n)$ coefficient matrix, corresponding to the state vectors $m_{t-i}$ and $\tilde{m}_{t-i}$, $\Phi:=[\Phi_1:\dots:\Phi_q]$ is $(n\times nq)$ matrix, $\mathsf{M}_{t-1}:=(m_{t-1}',\dots,m_{t-q}')'$ is an $(nq\times 1)$ state vector, and $\tilde{\mathsf{M}}_{t-1}:=(\tilde{m}_{t-1}',\dots,\tilde{m}_{t-q}')'$ is an $(nq\times 1)$ state vector, and $w_t$ is $(n\times 1)$ white noise process. Consequently, our models are given by the following systems
\begin{equation}\label{02098}
\begin{cases}
\Delta_t=-\text{diag}\{i_n+b_t\}m_t+\text{diag}\{i_n+A_0^k\psi_t\}m_{t-1}+u_t\\
x_t=\Pi^x\mathsf{X}_{t-1}+v_t\\
m_t=\Phi\mathsf{M}_{t-1}+\Pi^m\mathsf{X}_{t-1}+w_t
\end{cases}~~~\text{for}~t=1,\dots,T
\end{equation}
for the dividend--paying company, and
\begin{equation}\label{02099}
\begin{cases}
\tilde{b}_t=A_0^k\psi_t-\tilde{m}_t+\tilde{m}_{t-1}+u_t\\
x_t=\Pi^x\mathsf{X}_{t-1}+v_t\\
\tilde{m}_t=\Phi\tilde{\mathsf{M}}_{t-1}+\Pi^m\mathsf{X}_{t-1}+w_t
\end{cases}~~~\text{for}~t=1,\dots,T
\end{equation}
for the non--dividend paying company. The systems \eqref{02098} and \eqref{02099} are more compactly written by
\begin{equation}\label{02100}
\begin{cases}
y_t=\Psi_t z_t+\varphi_t+\xi_t\\
z_t=Az_{t-1}+\Pi_*^m\mathsf{X}_{t-1}+\eta_t
\end{cases}~~~\text{for}~t=1,\dots,T,
\end{equation}
where for the dividend--paying company, $y_t:=(\Delta_t',x_t')'$ is an $(\tilde{n}\times 1)$ vector, which is consists of observed variables' vectors $\Delta_t$ and $x_t$, $z_t:=\mathsf{M}_t$ is a ($nq\times 1$) state vector of the price--to--book ratios at times $t,\dots,t-q+1$, 
\begin{equation}\label{02101}
\Psi_t:=\begin{bmatrix}
-\text{diag}\{i_n+b_t\} & \text{diag}\{i_n+A_0^k\psi_t\} & 0 & \dots & 0\\
0 & 0 & 0 & \dots & 0
\end{bmatrix}
\end{equation}
is an ($\tilde{n}\times nq$) matrix, and $\varphi_t:=0$ and for the non--dividend paying company, $y_t:=(\tilde{b}_t',x_t')'$ is an $(\tilde{n}\times 1)$ vector, which is consists of observed variables' vectors $\tilde{b}_t$ and $x_t$, $z_t:=\tilde{\mathsf{M}}_t$ is a ($nq\times 1$) state vector of the log price--to--book ratios at times $t,\dots,t-q+1$, 
\begin{equation}\label{02102}
\Psi_t:=\begin{bmatrix}
-I_n & I_n & 0 & \dots & 0\\
0 & 0 & 0 & \dots & 0
\end{bmatrix}
\end{equation}
is an ($\tilde{n}\times nq$) matrix, and $\varphi_t:=((A_0^k\psi_t)',0)'$ is an $(\tilde{n}\times 1)$ vector, and $\xi_t=(u_t',v_t')'$ is an $(\tilde{n}\times 1)$ white noise process, $\eta_t:=(v_t',0,\dots,0)'$ is an ($nq\times 1$) random vector, $\Pi_*^m:=[(\Pi^m)':0:\dots:0]'$ is an $(nq\times n)$ matrix, whose first block is $\Pi^m$ and other blocks are zero, and
\begin{equation}\label{02103}
A:=\begin{bmatrix}
\Phi_1 & \dots & \Phi_{q-1} & \Phi_q\\
I_n & \dots & 0 & 0\\
\vdots & \ddots & \vdots & \vdots\\
0 & \dots & I_n & 0
\end{bmatrix}
\end{equation}
is an ($nq\times nq$) matrix. 

The stochastic properties of systems \eqref{02098}--\eqref{02100} are governed by the random variables $u_1,\dots,u_T,$ $v_1,\dots,v_T$, $w_1,\dots,w_T$, and $z_0$. We assume that the error random variables $u_t$ and $v_t$ for $t=1,\dots,T$ and initial book--to--price ratio $m_0$ or log book--to--price ratio $\tilde{m}_0$ are mutually independent, and follow normal distributions, namely,
\begin{equation}\label{02104}
z_0\sim \mathcal{N}(\mu_0,\Sigma_0), ~~~\xi_t\sim \mathcal{N}(0,\Sigma_{\xi\xi}),~~~w_t\sim \mathcal{N}(0,\Sigma_{ww}), ~~~\text{for}~t=1,\dots,T,
\end{equation}
where 
\begin{equation}\label{02105}
\Sigma_{\xi\xi}=\begin{bmatrix}
\Sigma_{uu} & \Sigma_{uv}\\
\Sigma_{vu} & \Sigma_{vv}
\end{bmatrix}
\end{equation}
is an $(\tilde{n}\times \tilde{n})$ covariance matrix of random error vector $\xi_t$.

For the rest of the subsection, we review the Kalman filtering for our model, see also \citeA{Hamilton94} and \citeA{Lutkepohl05}. For $t=0,\dots,T$, let $c_t:=(y_t',z_t')'$ be a $([\tilde{n}+nq]\times 1)$ vector, composed of the endogenous variable $y_t$ and the state vector $z_t$, and $\mathcal{F}_t:=(\mathcal{F}_0,\Delta_1',\dots,\Delta_t',x_1',\dots,x_t')$ and $\mathcal{F}_t:=(\mathcal{F}_0,\tilde{b}_1',\dots,\tilde{b}_t',x_1',\dots,x_t')$ be a available information at time $t$ of dividend--paying and non--dividend paying companies, respectively, where $\mathcal{F}_0:=(B_0',b_1',\dots,b_T',\psi_1',\dots,\psi_T')$ is an initial information for dividend--paying companies and $\mathcal{F}_0:=(B_0',\psi_1',\dots,\psi_T')$ is an initial information for non--dividend paying companies. Then, system \eqref{02100} can be written in the following form, which only depends on $z_{t-1}$
\begin{equation}\label{02106}
q_t=\begin{bmatrix}
y_t \\ z_t
\end{bmatrix}=\begin{bmatrix}
\Psi_t\Pi_*^m\mathsf{X}_{t-1}+\varphi_t\\ \Pi_*^m\mathsf{X}_{t-1}
\end{bmatrix}+\begin{bmatrix}
\Psi_t A \\ A
\end{bmatrix}z_{t-1}+\begin{bmatrix}
I_{\tilde{n}} & \Psi_t\\
0 & I_{nq}
\end{bmatrix}
\begin{bmatrix}
\xi_t \\ \eta_t
\end{bmatrix}~~~\text{for}~t=1,2,\dots.
\end{equation}
Because an error random vector $\zeta_t:=(\xi_t',\eta_t')'$ is independent of the information $\mathcal{F}_{t-1}$, conditional on $\mathcal{F}_{t-1}$, an expectation of a random vector $x_t:=(y_t,z_t)'$ is obtained by
\begin{equation}\label{02107}
\begin{bmatrix}
y_{t|t-1}\\z_{t|t-1}
\end{bmatrix}:=\begin{bmatrix}
\Psi_t\Pi_*^m\mathsf{X}_{t-1}+\varphi_t\\ \Pi_*^m\mathsf{X}_{t-1}
\end{bmatrix}+\begin{bmatrix}
\Psi_t A \\ A
\end{bmatrix}z_{t-1|t-1}
\end{equation}
for $t=1,\dots,T$, where $z_{0|0}:=(\mu_0,\dots,\mu_0)'$ is an ($nq\times 1$) initial value, which is consists of the vector $\mu_0$. If we use the tower property of conditional expectation and the fact that error random variables $\xi_t$ and $w_t$ are independent, and an error random vector $\zeta_t=(\xi_t',\eta_t')'$ is independent of the information $\mathcal{F}_{t-1}$, then it is clear that
\begin{equation}\label{02108}
\mathbb{E}\big((z_{t-1}-z_{t-1|t-1})\zeta_t'|\mathcal{F}_{t-1}\big)=0,~~~\mathbb{E}(\xi_t\eta_t'|\mathcal{F}_{t-1})=0,
\end{equation} 
for $t=1,\dots,T$. Consequently, it follows from equation \eqref{02106} that conditional on $\mathcal{F}_{t-1}$, a covariance matrix of the random vector $q_t$ is given by
\begin{equation}\label{02109}
\Sigma(q_t|t-1):=\text{Cov}(q_t|\mathcal{F}_{t-1})=\begin{bmatrix}
\Sigma(y_t|t-1) & \Sigma(z_t,y_t|t-1)'\\
\Sigma(z_t,y_t|t-1) & \Sigma(z_t|t-1)
\end{bmatrix}
\end{equation}
for $t=1,\dots,T$, where conditional on $\mathcal{F}_{t-1}$, a covariance matrix of the state vector $z_t$ is
\begin{equation}\label{02110}
\Sigma(z_{t}|t-1)=A\Sigma(z_{t-1}|t-1)A'+\Sigma_{\eta\eta}
\end{equation}
with $\Sigma_{\eta\eta}:=\text{Cov}(\eta_t)=\text{diag}\{\Sigma_{ww},0\}$ is an $(nq\times nq)$ matrix,
conditional on $\mathcal{F}_{t-1}$, a variance of the endogenous variable $y_t$ is
\begin{equation}\label{02111}
\Sigma(y_t|t-1):=\Psi_t\Sigma(z_{t}|t-1)\Psi_t'+\Sigma_{\xi\xi}
\end{equation}
with $\Sigma(z_{0}|0):=\text{diag}\{\Sigma_0,\dots,\Sigma_0\}$ is ($nq\times nq$) matrix, which is consists of the covariance matrix $\Sigma_0$, and conditional on $\mathcal{F}_{t-1}$, a covariance matrix between the endogenous variable $y_t$ and the state vector $z_t$ is
\begin{equation}\label{02112}
\Sigma(z_{t},y_t|t-1)=\Sigma(z_{t}|t-1)\Psi_t'
\end{equation}
As a result, due to equations \eqref{02107} and \eqref{02110}--\eqref{02112}, for given $\mathcal{F}_{t-1}$, a conditional distribution of the process $q_t$ is given by
\begin{equation}\label{02113}
q_t=\begin{bmatrix}
y_t \\ z_t
\end{bmatrix}
~\bigg|~\mathcal{F}_{t-1}\sim \mathcal{N}\left(\begin{bmatrix}
y_{t|t-1} \\ z_{t|t-1}
\end{bmatrix}, \begin{bmatrix}
\Sigma(y_t|t-1) & \Sigma(z_t,y_t|t-1)'\\
\Sigma(z_t,y_t|t-1) & \Sigma(z_t|t-1)
\end{bmatrix}
\right).
\end{equation}
It follows from the well--known formula of the conditional distribution of multivariate random vector and equation \eqref{02113} that a conditional distribution of the state vector $z_t$ given the endogenous variable $y_t$ and the information $\mathcal{F}_{t-1}$ is given by
\begin{equation}\label{02114}
z_t~|~y_t,\mathcal{F}_{t-1}\sim \mathcal{N}\Big(z_{t|t-1}+\mathcal{K}_{t}\big(y_t-y_{t|t-1}\big),\Sigma(z_t|t-1)-\mathcal{K}_{t}\Sigma(y_t|t-1)\mathcal{K}_t'\Big)
\end{equation}
for $t=1,\dots,T$, where $\mathcal{K}_t:=\Sigma(z_t,y_t|t-1)\Sigma^{-1}(y_t|t-1)$ is the Kalman filter gain. Therefore, since $\mathcal{F}_t=\{y_t,\mathcal{F}_{t-1}\}$, we have
\begin{equation}\label{02115}
z_{t|t}:=\mathbb{E}(z_t|\mathcal{F}_t)=z_{t|t-1}+\mathcal{K}_{t}\big(y_t-y_{t|t-1}\big),~~~t=1,\dots,T
\end{equation}
and
\begin{equation}\label{02116}
\Sigma(z_t|t):=\text{Cov}(z_t|\mathcal{F}_t)=\Sigma(z_t|t-1)-\mathcal{K}_{t}\Sigma(y_t|t-1)\mathcal{K}_t',~~~t=1,\dots,T.
\end{equation}

Because the error random vector $\zeta_t=(\xi_t,\eta_t)'$ for $t=T+1,T+2,\dots$ is independent of the full information $\mathcal{F}_T$ and the state vector at time $t-1$, $z_{t-1}$, it follows from equation \eqref{02100} and the tower property of conditional expectation that Kalman filter's forecast step is given by the following equations 
\begin{equation}\label{02117}
\begin{bmatrix}
y_{t|T}\\z_{t|T}
\end{bmatrix}=\begin{bmatrix}
\Psi_tz_{t|T}+\varphi_t\\ Az_{t-1|T}+\Pi_*^m\mathsf{X}_{t-1}
\end{bmatrix}~\text{and}~
\begin{bmatrix}
\Sigma(y_t|T)\\\Sigma(z_t|T)
\end{bmatrix}=\begin{bmatrix}
\Psi_t\Sigma(z_t|T)\Psi_t'+\Sigma_{\xi\xi} \\ A\Sigma(z_{t-1}|T)A'+\Sigma_{\eta\eta}
\end{bmatrix},~t=T+1,T+2,\dots.
\end{equation}

The Kalman filtering, which is considered the above provides an algorithm for filtering for the state vector $z_t$, which is the unobserved variable. To estimate the parameters of our models \eqref{02098} and \eqref{02099}, in addition to the Kalman filter, we also need to make inferences about the state vector $z_t$ for $t=1,\dots,T$ based on the full information $\mathcal{F}_T$, see below. Such an inference is called the smoothed estimate of the state vector $z_t$. The rest of the section is devoted to developing an algorithm, which is used to calculate the smoothed estimate $z_{t|T}:=\mathbb{E}(z_t|\mathcal{F}_T)$ for $t=0,\dots,T-1$.

Conditional on the information $\mathcal{F}_{t+1}$, a conditional distribution of a random vector $(z_t,z_{t+1})'$ is given by
\begin{equation}\label{02118}
\begin{bmatrix}
z_{t+1} \\ z_t
\end{bmatrix}~\bigg|~\mathcal{F}_{t}\sim \mathcal{N}\left(
\begin{bmatrix}
z_{t+1|t} \\ z_{t|t}
\end{bmatrix},
\begin{bmatrix}
\Sigma(z_{t+1}|t) & \Sigma(z_t,z_{t+1}|t)'\\
\Sigma(z_t,z_{t+1}|t) & \Sigma(z_{t}|t)
\end{bmatrix}\right)
\end{equation}
for $t=0,\dots,T-1$, where $\Sigma(z_t,z_{t+1}|t):=\text{Cov}(z_t,z_{t+1}|\mathcal{F}_t)$ is a covariance between state vectors at times $t$ and $t+1$ given the information $\mathcal{F}_{t}$. It follows from equation \eqref{02100} that the covariance is calculated by $\Sigma(z_t,z_{t+1}|t)=\Sigma(z_t|t)A'$. If we use the well--known formula of the conditional distribution of multivariate random vector once again, then a conditional distribution of the random state vector at time $t$ given the state at time $t+1$ and the information $\mathcal{F}_{t}$ is given by
\begin{equation}\label{02119}
z_t~|~z_{t+1},\mathcal{F}_{t}\sim \mathcal{N}\Big(z_{t|t}+\mathcal{S}_{t}\big(z_{t+1}-z_{t+1|t}\big),\Sigma(z_t|t)-\mathcal{S}_t\Sigma(z_{t+1}|t)\mathcal{S}_t'\Big)
\end{equation}
for $t=0,\dots,T-1$, where $\mathcal{S}_{t}:=\Sigma(z_t,z_{t+1}|t)\Sigma^{-1}(z_{t+1}|t)$ is the Kalman smoother gain. Because conditional on the state vector $z_{t+1}$, the state vector at time $t$, $z_t$, is independent of an endogenous variable vector $(y_{t+1},\dots,y_T)'$, for each $t=0,\dots,T-1$, it holds $\mathbb{E}(z_t|z_{t+1},\mathcal{F}_{T})=\mathbb{E}(z_t|z_{t+1},\mathcal{F}_t)=z_{t|t}+\mathcal{S}_{t}\big(z_{t+1}-z_{t+1|t}\big)$. Therefore, it follows from the tower property of the conditional expectation and conditional expectation in equation \eqref{02119} that the smoothed inference of the state vector $z_t$ is obtained by
\begin{equation}\label{02120}
z_{t|T}=\mathbb{E}\big(\mathbb{E}(z_t|z_{t+1},\mathcal{F}_T)\big|\mathcal{F}_T\big)=z_{t|t}+\mathcal{S}_{t}\big(z_{t+1|T}-z_{t+1|t}\big)
\end{equation}
for $t=0,\dots,T-1$. Using equation \eqref{02120} a difference between the state vector $z_t$ and its Kalman smoother $z_{t|T}$ is represented by
\begin{equation}\label{02121}
z_t-z_{t|T}=z_t-\big[z_{t|t}+\mathcal{S}_{t}(z_{t+1}-z_{t+1|t})\big]+\mathcal{S}_{t}(z_{t+1}-z_{t+1|T}).
\end{equation}
Observe that the square bracket term in the above equation is the conditional expectation of the state vector at time $t$, which is given in equation \eqref{02119}. Thus, if we use the conditional covariance matrix of the state vector $z_t$, which is given in equation \eqref{02119} and use the tower property of conditional expectation once more, then we obtain that
\begin{equation}\label{02122}
\Sigma(z_t|T)=\mathbb{E}\big((z_t-z_{t|T})(z_t-z_{t|T})'\big|\mathcal{F}_T\big)=\Sigma(z_t|t)-\mathcal{S}_{t}\big(\Sigma(z_{t+1}|t)-\Sigma(z_{t+1}|T)\big)\mathcal{S}_t'
\end{equation}
and
\begin{equation}\label{02123}
\Sigma(z_{t},z_{t+1}|T)=\mathbb{E}\big((z_{t}-z_{t|T})(z_{t+1}-z_{t+1|T})'\big|\mathcal{F}_T\big)=\mathcal{S}_{t}\Sigma(z_{t+1}|T)
\end{equation}
for $t=0,\dots,T-1$. 

Firstly, let us consider the dividend--paying firm, which is given by system \eqref{02098}. In the EM algorithm, one considers a joint density function of a random vector, which is composed of observed variables and state (latent) variables. In our cases, the vectors of observed variables and state variables correspond to the vector of dividend--to--book ratios and economic variables, $y:=(y_1',\dots,y_T')'$, and a vector of price--to--book ratio vectors, $m:=(m_0',\dots,m_T')'$, respectively. Interesting usages of the EM algorithm in econometrics can be found in \citeA{Hamilton90} and \citeA{Schneider92}. Let us denote the joint density function by $f_{\Delta,m}(\Delta,m)$. The EM algorithm consists of two steps. 

In the expectation (E) step of the EM algorithm, one has to determine the form of an expectation of log of the joint density given the full information $\mathcal{F}_T$. We denote the expectation by $\Lambda(\theta|\mathcal{F}_T)$, that is, $\Lambda(\theta|\mathcal{F}_T):=\mathbb{E}\big(\ln(f_{y,m}(y,m))|\mathcal{F}_T\big)$. For our model \eqref{02098}, one can show that the expectation of log of the joint density of the vectors of the observed variables $y$ and the vector of the price--to--book ratio vectors $m$ is
\begin{eqnarray}\label{02124}
\Lambda(\theta|\mathcal{F}_T)&=&-\frac{(\tilde{n}+n+1)T}{2}\ln(2\pi)-\frac{T}{2}\ln(|\Sigma_{\xi\xi}|)-\frac{T}{2}\ln(|\Sigma_{ww}|)-\frac{1}{2}\ln(|\Sigma_{0}|)\nonumber\\
&-&\frac{1}{2}\sum_{t=1}^T\mathbb{E}\Big[u_t'\Omega_{uu}u_t\Big|\mathcal{F}_T\Big]-\sum_{t=1}^T\mathbb{E}\Big[u_t'\Omega_{uv}v_t\Big|\mathcal{F}_T\Big]-\frac{1}{2}\sum_{t=1}^T\mathbb{E}\Big[v_t'\Omega_{vv} v_t\Big|\mathcal{F}_T\Big]\\
&-&\frac{1}{2}\sum_{t=1}^T\mathbb{E}\Big[w_t'\Sigma_{ww}^{-1} w_t\Big|\mathcal{F}_T\Big]-\frac{1}{2}\mathbb{E}\Big[\big(\tilde{m}_0-\mu_0\big)'\Sigma_0^{-1}\big(\tilde{m}_0-\mu_0\big)\Big|\mathcal{F}_T\Big],\nonumber
\end{eqnarray}
where $\theta:=\big(\text{vec}(A_0^k)',\mu_0',\text{vec}(\Phi)',\text{vec}(\Pi^x)',\text{vec}(\Pi^m)',\text{vech}(\Sigma_{\xi\xi})',\text{vech}(\Sigma_{ww})',\text{vech}(\Sigma_0)'\big)'$ is a $([n(l+q+nq)+\tilde{n}(l+\tilde{n}p)+(\tilde{n}(\tilde{n}+1)+n(n+1)+nq(nq+1))/2]\times 1)$ vector, which consists of all parameters of the model \eqref{02098}, $\Omega_{uu}$, $\Omega_{uv}$, $\Omega_{vu}$, and $\Omega_{vv}$ are the partitions of the matrix $\Sigma_{\xi\xi}^{-1}$, corresponding to the random vector $\xi_t=(u_t',v_t')'$, 
\begin{eqnarray}
u_t&=&\Delta_t+\text{diag}\{i_n+b_t\}m_t-\text{diag}\{i_n+A_0^k\psi_t\}m_{t-1}\label{02125}\\
&=&\Delta_t+M_t(i_n+b_t)-M_{t-1}(i_n+A_0^k\psi_t),\label{02126}
\end{eqnarray}
\begin{equation}\label{02127}
v_t=x_t-\Pi^x\mathsf{X}_{t-1},
\end{equation}
and
\begin{equation}\label{02128}
w_t=m_t-\Phi z_{t-1}-\Pi^m\mathsf{X}_{t-1}
\end{equation}
are the $(n\times 1)$, $(\ell\times 1)$, $(n\times 1)$ white noise processes, respectively, and $M_t:=\text{diag}\{m_t\}$ is an $(n\times n)$ diagonal matrix, whose diagonal elements are $m_t$. 

In the maximization (M) step of the EM algorithm, we need to find a maximum likelihood estimator $\hat{\theta}$ that maximizes the expectation, which is determined in the E step. According to equation \eqref{02126}, the white noise process $u_t$ can be written by
\begin{equation}\label{02129}
u_t=\Delta_t+M_t(i_n+b_t)-M_{t-1}i_n-(\psi_t'\otimes M_{t-1})a_0^k
\end{equation}
where $a_0^k:=\text{vec}(A_0^k)$ is a vectorization of the matrix $A_0^k$. As a result, a partial derivative of the log--likelihood function with respect to the parameter $a_0^k$ is given by
\begin{equation}\label{02130}
\frac{\partial \Lambda(\theta|\mathcal{F}_T)}{\partial (a_0^k)'}=\sum_{t=1}^T\mathbb{E}\bigg[\Big(u_t'\Omega_{uu}+v_t'\Omega_{vu}\Big)(\psi_t'\otimes M_{t-1})\bigg|\mathcal{F}_T\bigg].
\end{equation}
Let $J_m:=[I_n:0:\dots:0]$ be $(n\times nq)$ matrix, whose first block matrix is $I_n$ and other blocks are zero, $z_{t-1,t-1|T}:=\mathbb{E}\big[z_{t-1}z_{t-1}'\big|\mathcal{F}_T\big]=\Sigma(z_{t-1}|T)+z_{t-1|T}z_{t-1|T}'$ be an $(nq\times nq)$ smoothed matrix, and $z_{t-1,t|T}:=\mathbb{E}\big[z_{t-1}z_t'\big|\mathcal{F}_T\big]=\mathcal{S}_{t-1}\Sigma(z_{t-1}|T)+z_{t-1|T}z_{t|T}'$ be an $(nq\times nq)$ smoothed matrix, see equation \eqref{02123}. The matrix $J_m$ can be used to extract the smoothed inference vector $m_{t|T}$ and smoothed inference matrices $m_{t-1,t-1|T}:=\mathbb{E}\big[m_{t-1}m_{t-1}'\big|\mathcal{F}_T\big]$ and $m_{t-1,t|T}:=\mathbb{E}\big[m_{t-1}m_t'\big|\mathcal{F}_T\big]$ from the smoothed inference vector $z_{t|T}$ and smoothed inference matrices $z_{t-1,t-1|T}:=\mathbb{E}\big[z_{t-1}z_{t-1}'\big|\mathcal{F}_T\big]$ and $z_{t-1,t|T}:=\mathbb{E}\big[z_{t-1}z_t'\big|\mathcal{F}_T\big]$, that is, $m_{t|T}=J_mz_{t|T}$, $m_{t-1,t-1|T}=J_mz_{t-1,t-1|T}J_m'$, and $m_{t-1,t|T}=J_mz_{t-1,t|T}J_m'$.
Since for all $a,b\in\mathbb{R}^n$ vectors and $C\in\mathbb{R}^{n\times n}$ matrix, $\text{diag}\{a\}C\text{diag}\{b\}=C\odot(ab')$, it follows from above equation \eqref{02130} that a ML estimator of the parameter $a_0^k$ is obtained by the following equation
\begin{eqnarray}
\hat{a}_0^k&:=&\Bigg(\sum_{t=1}^T\Big(\psi_t\psi_t'\otimes \Big[\Omega_{uu}\odot \big(J_mz_{t-1,t-1|T}J_m'\big)\Big]\Big)\Bigg)^{-1}\nonumber\\
&\times&\sum_{t=1}^T\bigg(\psi_t\otimes \Big[\text{diag}\big\{J_mz_{t-1|T}\big\}\Big(\Omega_{uu}\Delta_t+\Omega_{uv}(x_t-\Pi^x\mathsf{X}_{t-1})\Big)\label{02131}\\
&+&\Big(\Omega_{uu}\odot \big(J_mz_{t-1,t|T}\big)\Big)\big(i_n+b_t\big)-\Big(\Omega_{uu}\odot \big(J_mz_{t-1,t-1|T}\big)\Big)i_n\Big\}\bigg).\nonumber
\end{eqnarray}

Due to equation \eqref{02127}, white noise process $v_t$ is represented by $v_t=x_t-(\mathsf{X}_{t-1}'\otimes I_\ell)\pi^x$, where $\pi^x:=\text{vec}(\Pi^x)$ is a vectorization of the matrix $\Pi^x$. Consequently, a partial derivative of the log--likelihood function with respect to the parameter $\pi^x$ is given by
\begin{equation}\label{02132}
\frac{\partial \Lambda(\theta|\mathcal{F}_T)}{\partial (\pi^x)'}=\sum_{t=1}^T\mathbb{E}\bigg[\Big(v_t'\Omega_{vv}+u_t'\Omega_{uv}\Big)(\mathsf{X}_{t-1}'\otimes I_\ell)\bigg|\mathcal{F}_T\bigg].
\end{equation}
Let $\bar{z}_{\bullet|T}:=[z_{1|T}:\dots:z_{T|T}]$ be an $(nq\times T)$ smoothed inference matrix and $\bar{z}_{-1|T}:=[z_{0|T}:\dots:z_{T-1|T}]$ be an $(nq\times T)$ smoothed inference matrix, which is backshifted the matrix $\bar{z}_{\bullet|T}$ by one period. After some manipulation, we obtain an ML estimator of the parameter $\Pi^x$
\begin{eqnarray}
\hat{\Pi}^x&:=&\Big(\bar{x}+\Omega_{vv}^{-1}\Omega_{vu}\big(\bar{\Delta}+(J_m\bar{z}_{\bullet|T})\odot(i_n\otimes i_T'+\bar{b})\label{02133}\\
&-&(J_m\bar{z}_{-1|T})\odot(i_n\otimes i_T'+A_0^k\bar{\psi})\big)\Big)\bar{\mathsf{X}}'(\bar{\mathsf{X}}\bar{\mathsf{X}}')^{-1}.\nonumber
\end{eqnarray}

Because of equation \eqref{02128}, white noise process $w_t$ is represented by $w_t=J_mz_t-\Phi z_{t-1}- (\mathsf{X}_{t-1}'\otimes I_n)\pi^m$, where $\pi^m:=\text{vec}(\Pi^m)$ is a vectorization of the matrix $\Pi^m$. Therefore, a partial derivative of the log--likelihood function with respect to the parameter $\pi^m$ is given by
\begin{equation}\label{02134}
\frac{\partial \Lambda(\theta|\mathcal{F}_T)}{\partial (\pi^m)'}=\sum_{t=1}^T\mathbb{E}\Big[w_t'\Omega_{ww}(\mathsf{X}_{t-1}'\otimes I_n)\bigg|\mathcal{F}_T\Big].
\end{equation}
From the above equation, it can be shown that an ML estimator of the parameter $\Pi^m$ is given by
\begin{equation}\label{02135}
\hat{\Pi}^m:=\big(J_m\bar{z}_{\bullet|T}-\Phi\bar{z}_{-1|T}\big)\bar{\mathsf{X}}'(\bar{\mathsf{X}}\bar{\mathsf{X}}')^{-1}.
\end{equation}

According to equation \eqref{02128}, white noise process $w_t$ is represented by $w_t=J_mz_t-(z_{t-1}'\otimes  I_n)\phi-\Pi^m\mathsf{X}_{t-1}$, where $\phi:=\text{vec}(\Phi)$ is a vectorization of the matrix $\Phi$. Therefore, a partial derivative of the log--likelihood function with respect to the parameter $\phi$ is given by
\begin{equation}\label{02136}
\frac{\partial \Lambda(\theta|\mathcal{F}_T)}{\partial \phi'}=\sum_{t=1}^T\mathbb{E}\Big[w_t'\Omega_{ww}(z_{t-1}'\otimes I_n)\bigg|\mathcal{F}_T\Big].
\end{equation}
Since $(z_{t-1}\otimes I_n)\Sigma_{ww}^{-1}J_mz_t=\text{vec}(\Sigma_{ww}'J_mz_tz_{t-1}')=(z_{t-1}z_t'J_m'\otimes \Sigma_{ww}^{-1})\text{vec}(I_n)$, after some manipulation, we arrive at an ML estimator of the parameter $\Phi$
\begin{equation}\label{02137}
\hat{\Phi}:=\Bigg(\sum_{t=1}^TJ_mz_{t-1,t|T}'-\Pi^m\bar{\mathsf{X}}z_{-1|T}'\Bigg)\Bigg(\sum_{t=1}^Tz_{t-1,t-1|T}\Bigg)^{-1}.
\end{equation}
where $J_m:=[I_n:0]$ is an $(n\times nq)$ matrix, whose first block matrix is $I_n$ and other blocks equal zero.

For estimators of the covariance matrices $\Sigma_{\eta\eta}$, $\Sigma_{ww}$, and $\Sigma_0$, the following formulas holds
\begin{equation}\label{02138}
\hat{\Sigma}_{\xi\xi}:=\frac{1}{T}\sum_{t=1}^T\mathbb{E}[\xi_t\xi_t'|\mathcal{F}_T]=\frac{1}{T}\sum_{t=1}^T\begin{bmatrix}
\mathbb{E}[u_tu_t'|\mathcal{F}_T] & \mathbb{E}[u_tv_t'|\mathcal{F}_T]\\
\mathbb{E}[v_tu_t'|\mathcal{F}_T] & \mathbb{E}[v_tv_t'|\mathcal{F}_T]
\end{bmatrix},
\end{equation}
\begin{equation}\label{02139}
\hat{\Sigma}_{ww}:=\frac{1}{T}\sum_{t=1}^T\mathbb{E}[w_tw_t'|\mathcal{F}_T], 
~~~~\text{and}~~~\hat{\Sigma}_0:=\Sigma(z_0|T).
\end{equation}
To calculate the conditional expectations $\mathbb{E}\big(\xi_t\xi_t'|\mathcal{F}_T\big)$ and $\mathbb{E}\big(w_tw_t'|\mathcal{F}_T\big)$, observe that the random error processes at time $t$ of the log book value growth rate process and the log multiplier process can be represented by
\begin{eqnarray}\label{02140}
u_t&=&u_{t|T}-\Psi_{\Delta,t}(z_t-z_{t|T})\nonumber\\
v_t&=&x_t-\Pi^x\mathsf{X}_{t-1}\\
w_t&=&w_{t|T}+J_m(z_t-z_{t|T})-\Phi(z_{t-1}-z_{t-1|T})\nonumber.
\end{eqnarray}
where $\Psi_{\Delta,t}:=[-\text{diag}\{i_n+b_t\}:\text{diag}\{i_n+A_0^k\psi_t\}:0:\dots:0]$ is an $(n\times nq)$ matrix and its third to $q$--th block matrices are zero. Therefore, as $v_t$, $u_{t|T}$, and $v_{t|T}$ are measurable with respect to the full information $\mathcal{F}_T$ (known at time $T$), it follows from equations \eqref{02140} that 
\begin{equation}\label{02141}
\mathbb{E}\big(u_tu_t'|\mathcal{F}_T\big)=u_{t|T}u_{t|T}'+\Psi_{\Delta,t}\Sigma(z_t|T)\Psi_{\Delta,t}',~~~
\mathbb{E}\big(u_tv_t'|\mathcal{F}_T\big)=u_{t|T}v_t',~~~
\mathbb{E}\big(v_tv_t'|\mathcal{F}_T\big)=v_tv_t',
\end{equation}
and
\begin{eqnarray}\label{02142}
\mathbb{E}\big(w_tw_t'|\mathcal{F}_T\big)&=&w_{t|T}w_{t|T}'+J_m\Sigma(z_t|T)J_m'+\Phi\Sigma(z_{t-1}|T)\Phi'\\ 
&-&J_m\Sigma(z_t|T)\mathcal{S}_{t-1}'\Phi'-\Phi\mathcal{S}_{t-1}\Sigma(z_t|T)J_m'.\nonumber
\end{eqnarray}

If we substitute equation \eqref{02141} into \eqref{02138}, then under suitable conditions the zig--zag iteration that corresponds to equations \eqref{02107}, \eqref{02110}, \eqref{02111}, \eqref{02115}, \eqref{02116}, \eqref{02120}, \eqref{02122}, \eqref{02131}, \eqref{02133}, \eqref{02135}, and \eqref{02137}--\eqref{02139} converges to the maximum likelihood estimators of our log private company valuation model. 

Now we consider the non--dividend paying companies. For the non--dividend paying companies, their white noise process $u_t$ is given by
\begin{equation}\label{02144}
u_t=\tilde{b}_t-A_0^k\psi_t+J_m(z_t-z_{t-1}).
\end{equation}
In a similar manner as the public companies, it can be shown that ML estimators of the parameters $A_0^k$ and $\Pi^x$ are obtained by the following equations
\begin{equation}\label{02145}
\hat{A}_0^k:=\Big(\bar{b}+J_m\big(\bar{z}_{\bullet|T}-\bar{z}_{-1|T}\big)+\Omega_{uu}^{-1}\Omega_{uv}\big(\bar{x}-\Pi^x\bar{\mathsf{X}}\big)\Big)\bar{\psi}'(\bar{\psi}\bar{\psi}')^{-1}
\end{equation}
and
\begin{equation}\label{02146}
\hat{\Pi}^x:=\Big(\bar{x}+\Omega_{vv}^{-1}\Omega_{vu}\big(\bar{b}-A_0^k\bar{\psi}+J_m\big(\bar{z}_{\bullet|T}-\bar{z}_{-1|T}\big)\big)\Big)\bar{\mathsf{X}}'(\bar{\mathsf{X}}\bar{\mathsf{X}}')^{-1}.
\end{equation}
Since $u_t=u_{t|T}-\Psi_{\tilde{b},t}(z_t-z_{t|T})$, where $u_{t|T}=\tilde{b}_t-A_0^k\psi_t+J_m(z_{t|T}-z_{t-1|T})$ is an $(n\times 1)$ smoothed white noise process and $\Psi_{\tilde{b},t}:=[-I_n:I_n:0:\dots:0]$ is an $(n\times nq)$ matrix, whose third to $q$--th block matrices are zero, one finds that
\begin{equation}\label{02147}
\mathbb{E}\big(u_tu_t'|\mathcal{F}_T\big)=u_{t|T}u_{t|T}'+\Psi_{\tilde{b},t}\Sigma(z_t|T)\Psi_{\tilde{b},t}'.
\end{equation}
Using the same method as the dividend--paying company, one can obtain other ML estimators. The other ML estimators of parameters of the non--dividend paying company are given by equations \eqref{02145}--\eqref{02147}. 

Let us suppose that the parameter estimation of our model is obtained. Then, a smoothed inference of the market value process at time $t$ of the private company is calculated by the following formula
\begin{equation}\label{02143}
V_{t|T}=m_{t|T}B_t, ~~~t=0,1,\dots,T,
\end{equation}
where $m_{t|T}=\exp\{\tilde{m}_{t|T}\}$ is a smoothed multiplier vector at time $t$. Also, an analyst can forecast the market value process of the private company by using equations \eqref{02117} and \eqref{02143}.

\section{Numerical Results}

We start by applying the estimation method for parameter estimation of our model, see Section 3.4. For means of illustration, we have chosen three companies from different sectors (Healthcare, Financial Services, and Consumer), listed in the S\&P 500 index. In order to increase the number of price and dividend observation points, we take quarterly data instead of yearly data. Our data covers a period from Q1 1990 to Q3 2021. That leads to $T=127$ observations for Johnson \& Johnson, PepsiCo, and JPMorgan. All quarterly price and dividend data have been collected from Thomson Reuters Eikon.  

The dividends of the selected companies have different patterns. In particular, JPMorgan cut its dividend by a huge amount due to the 2008/2009 financial crises, and the other companies have continuously increasing dividend dynamics, which are not affected by the 2008/2009 financial crises. For our model, we assume for all companies, that a default never occurs. 

We present estimations of the parameters for the selected companies in Table 1. The 2--9th rows of Table 1 correspond to the required rate of returns of the companies modeled by the regime--switching process with three regimes and the 10--13th rows of the same Table correspond to the required rate of returns of the companies take constant values (the regime--switching process takes one regime).

In order to obtain estimations of the parameters, which correspond to the 2--9th rows of Table 1 we assume that the regime--switching process $s_t$ follows a Markov chain with three regimes, namely, up regime (regime 1), normal regime (regime 2), and down regime (regime 3) and we use equations \eqref{02021}--\eqref{02025}. Since explanations are comparable for the other companies, we will give explanations only for PepsiCo. In the 2nd row of Table 1, we provide estimations of the parameters $\tilde{k}(1),\tilde{k}(2),\tilde{k}(3)$. For PepsiCo, in regimes 1, 2, and 3, estimations of the required rate of return are 19.44\%, 3.37\%, and --20.86\%, respectively. For example, in the normal regime, the required rate of return of PepsiCo could be 2.89\% on average. 

The 3--5th rows of Table 1 correspond to the transition probability matrix $P$. For the selected companies, their transition probability matrices $P$s are ergodic, where ergodic means that one of the eigenvalues of $P$ is unity and that all other eigenvalues of $P$ are inside the unit circle, see \citeA{Hamilton94}. From the 3rd row of Table 1 one can deduce that if the required rate of return of PepsiCo is in the up regime then in the next period, it will switch to the normal regime with a probability of 0.814 or the down regime with a probability of 0.186 because it can not be in the up regime due to zero probability. If the required rate of return of PepsiCo is in the normal regime, corresponding to row 4 of the Table, then in the next period, it can not switch to the up regime because of zero probability, the normal regime with a probability of 0.962, or the down regime with a probability of 0.038. Finally, if the required rate of return of PepsiCo is in the down regime then in the next period, it will switch to the up regime with a probability of 0.840 or the down regime with a probability of 0.160 due to the normal regime's zero probability, see 5th row of the same Table. 

We provide the average persistence times of the regimes in the 6th row of Table 1. The average persistence time of the regime $j$ is defined by $\tau_j:=1/(1-p_{jj})$ for $j=1,2,3.$ From Table 1, one can conclude that up, normal, and down regimes of PepsiCo's required rate of return will persist on average for 1.0, 25.6, and 1.3 quarters, respectively.

In the 7th row of Table 1, we give ergodic probabilities $\pi$ of the selected companies. Ergodic probability vector $\pi$ of an ergodic Markov chain is defined from an equation $P\pi=\pi$. The ergodic probability vector represents long--run probabilities, which do not depend on the initial probability vector $\rho=z_{1|0}$. After sufficiently long periods, the required rate of return of PepsiCo will be in the up regime with a probability of 0.042, the normal regime with a probability of 0.908, or the down regime with a probability of 0.050, which are irrelevant to initial regimes.

The 8th row of Table 1 is devoted to long--run expectations of the required rate of returns of the selected companies. The long--run expectation of the required rate of return is defined by $k_\infty:=\lim_{t\to \infty}\mathbb{E}(k(s_t)).$ For PepsiCo, it equals 2.83\%. So that after long periods, the average required rate of return of PepsiCo converges to 2.83\%.

\begin{table}[htbp]
  \centering
  \caption{ML Estimation for the Markov--Switching DDM}
  \scalebox{0.97}{    
  \begin{tabular}{|c|c|c|c|c|c|c|c|c|c|c|}
    \hline
    Row   & Prmtrs & \multicolumn{3}{c|}{Johnson \& Johnson} & \multicolumn{3}{c|}{PepsiCo} & \multicolumn{3}{c|}{JPMorgan} \\
    \hline
    2.     & $k(j)$  & 14.88\% & 3.30\% & --22.41\% & 19.44\% & 3.37\% & --20.86\% & 41.42\% & 2.79\% & --45.85\% \\
    \hline
    3.     & \multirow{3}{*}{$P$} & 0.000 & 1.000 & 0.000 & 0.000 & 0.814 & 0.186 & 0.193 & 0.807 & 0.000 \\
\cline{1-1}\cline{3-11}    4.     &       & 0.036 & 0.937 & 0.027 & 0.000 & 0.962 & 0.038 & 0.007 & 0.954 & 0.039 \\
\cline{1-1}\cline{3-11}    5.     &       & 0.756 & 0.000 & 0.244 & 0.840 & 0.000 & 0.160 & 1.000 & 0.000 & 0.000 \\
    \hline
    6.     & $\tau_j$   & 1.000 & 15.79 & 1.322 & 1.000 & 26.60 & 1.191 & 1.239 & 21.60 & 1.000 \\
    \hline
    7.     & $\pi$    & 0.058 & 0.910 & 0.033 & 0.042 & 0.908 & 0.050 & 0.052 & 0.912 & 0.035 \\
    \hline
    8.     & $\tilde{k}_\infty$     & \multicolumn{3}{c|}{3.12\%} & \multicolumn{3}{c|}{2.83\%} & \multicolumn{3}{c|}{3.09\%} \\
    \hline
    9.     & $\sigma_3$ & \multicolumn{3}{c|}{0.064} & \multicolumn{3}{c|}{0.070} & \multicolumn{3}{c|}{0.124} \\
    \hline
    10.    & $\tilde{k}$     & \multicolumn{3}{c|}{3.14\%} & \multicolumn{3}{c|}{2.84\%} & \multicolumn{3}{c|}{3.08\%} \\
    \hline
    11.    & $\tilde{k}_L$  & \multicolumn{3}{c|}{1.66\%} & \multicolumn{3}{c|}{1.18\%} & \multicolumn{3}{c|}{--0.06\%} \\
    \hline
    12.    & $\tilde{k}_U$  & \multicolumn{3}{c|}{4.62\%} & \multicolumn{3}{c|}{4.50\%} & \multicolumn{3}{c|}{6.23\%} \\
    \hline
    13.    & $\sigma_1$ & \multicolumn{3}{c|}{0.084} & \multicolumn{3}{c|}{0.094} & \multicolumn{3}{c|}{0.178} \\
    \hline
    \end{tabular}%
  \label{tab01}%
}
\end{table}%

In the 9th row of Table 1, we present parameter estimations of standard deviations of the error random variables $u_t$ for the selected companies. For PepsiCo, the parameter estimation equals 0.079. The 13th row of Table 1 corresponds to the parameter estimations of standard deviations, in which the required rate of returns of the companies are modeled by regime--switching process with one regime. For PepsiCo, the parameter estimation equals 0.094, where we used equation \eqref{eq08}. As we compare the 9th row and 13th row of the Table, we can see that the estimations that correspond to the regime--switching process with three regimes are lower than the ones that correspond to the regime--switching process with one regime. 

Finally, the log required rate of returns estimation at time Q3 2021 of the firms are presented in row ten of the Table, while the corresponding 95\% confidence intervals are included in rows 11 and 12 below. To calculate the log required rate of returns estimation and confidence bands, we used equations \eqref{02149}/\eqref{02150} and \eqref{02155}. The Table further illustrates average log returns (2.84\% for PepsiCo) and the return variability, as the return is supposed to lie within the (1.18\%, 4.50\%) interval with a 95\% probability. It is worth mentioning that as calculations are based on the log required rate of return, we should convert them to the required rate of return using the formula $k_i=\exp\{\tilde{k}_i\}-1$ for each company, see equation \eqref{02156}. In particular, for PepsiCo company, its point estimation of the required rate of return is $k_2=\exp\{2.84\%\}-1=2.88\%$ and 95\% confidence interval is $\big(\exp\{1.18\%\}-1,\exp\{4.50\%\}-1\big)=(1.18\%,4.60\%)$. Also, note that since the required rate of return estimation expresses the average quarterly return of the companies, we can convert them yearly based using the formula $(1+k)^4-1$. 

For the selected firms, it will be interesting to plot the probabilistic inferences with the log return series. For each period $t=1,\dots,T$ and each firm, the probabilistic inferences are calculated by equation \eqref{02021} and the log return series are calculated by the formula $\tilde{k}_t:=\ln\big((P_t+d_t)/P_{t-1}\big)$. In Figure 1, we plotted the resulting series as a function of period $t$. In Figure 1, the left axis corresponds to the return series, while the right axis corresponds to the probabilistic inference series for each company. 

\begin{figure}[!h]\label{Fig1}
\centering
\caption{Returns VS Regime Probabilities of Selected Companies} 
\includegraphics[width=165mm]{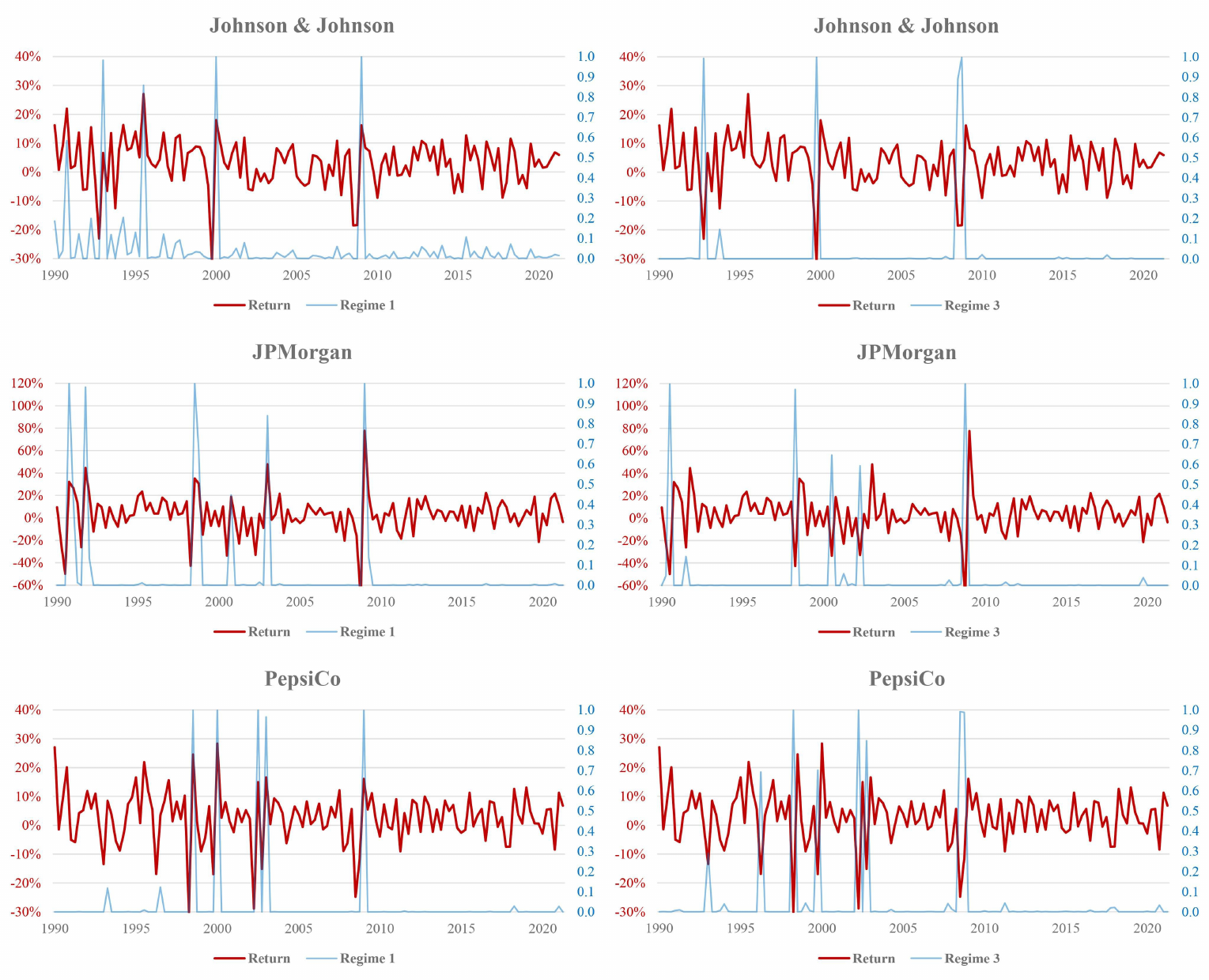} 
\end{figure}

From the Figure, and the 9th and 13th rows of Table 1, we can deduce that the regime--switching processes with three regimes are suited to explain the required rate of return series as compared to the regime--switching processes with one regime.  

From the Figure, we may expect that the log required rate of returns of the companies follow conditional heteroscedastic models. By using Eviews 12 the econometric program, one can conclude that the log required rate of returns of Johnson \& Johnson and PepsiCo, which are demeaned by intercepts are white noise processes, while the log required rate of return of JPMorgan can be modeled by AR(0)--ARCH(1) process, namely,
$$\tilde{k}_{3,t}=0.022+\xi_{3,t},~~~\xi_{3,t}=\sigma_{3,t}\varepsilon_{3,t}, ~~~\sigma_{3,t}^2=0.015+0.616\xi_{3,t}^2.$$
Because the coefficient of $\xi_{3,t}^2$ of the above equation lies in an interval $[0,1)$, the log required rate of return process of JPMorgan is covariance stationary process, see \citeA{McNeil05}.

Finally, let us consider the Bayesian estimator of the companies' log required rate of returns. Since each of the log required rate of return processes are covariance stationary, we take $\delta_i=0$ for $i=1,2,3$. By using Akaike's and Schwartz's information criterion, we deduce that for the three companies, an order of simple VAR$(p)$ process is $p=1$. For this reason, we choose an order of Bayesian VAR$(p)$ process by $p=1$. Observe that because of the fact that all companies' log required rate of returns are covariance stationary, the prior expectation matrix $\Pi_0$ equals zero, i.e., $\Pi_0=0$. Since each company's log required rate of return follows AR(0) process, for each $i=1,2,3$, we estimate the parameter $\sigma_i^2$ by the sample variance $s_i^2=\frac{1}{T}\sum_{t=1}^T(k_{i,t}-\bar{k}_i)^2$, where $\bar{k}_i=\frac{1}{T}\sum_{t=1}^Tk_{i,t}$ is the sample mean of $i$--th company's log required rate of return. To obtain the Bayesian estimator, we need to define the other hyperparameters: $\lambda_1=5^2$, $\lambda_2=0.2^2$, $\nu_0=\tilde{n}+2=5$, and $V_0=\text{diag}\{s_1^2,s_2^2,s_3^2\}$. By using equations \eqref{02057} and \eqref{02058}, we obtain the Bayesian estimators, which are given in Table 2 of the parameters $\Pi$ and $\Sigma$.

\begin{table}[htbp]
  \centering
  \caption{Bayesian Estimation of the Selected Companies}
    \begin{tabular}{|c|c|c|c|c|c|c|c|}
    \hline
          & \multicolumn{4}{c|}{$\Pi_{*|T}$}       & \multicolumn{3}{c|}{$\Sigma_{*|T}$} \\
    \hline
    Johnson \& Johnson & 0.0311 & --0.0034 & --0.0436 & 0.0175 & 0.0109 & 0.0061 & 0.0047 \\
    \hline
    PepsiCo & 0.0294 & 0.1246 & --0.2111 & --0.0320 & 0.0061 & 0.0117 & 0.0093 \\
    \hline
    JPMorgan & 0.0299 & 0.1205 & 0.1007 & --0.1968 & 0.0047 & 0.0093 & 0.0449 \\
    \hline
    \end{tabular}%
  \label{tab02}%
\end{table}%

It should be noted that by applying the results in Table 2 and the Gibbs sampling method, which is mentioned before in Section 2.2, one may make inferences about the parameters and forecasts of the log required rate of returns of the companies.

\section{Conclusion}

The most popular practical method, which is used to estimate the required rate of return on equity is the CAPM. However, the CAPM is sensitive to its inputs. Therefore, in this paper, instead of the traditional CAPM and its descendant versions, we introduce new estimation methods, covering the ML methods with regime--switching, the Bayesian method, and the Kalman filtering to estimate the required rate of return on equity. 

The required rate of return on equity has some practical applications. For example, in addition to its usage in stock valuation, it is an ingredient of the WACC. If a company is financed by liabilities, which are publicly traded in the exchanges, one can estimate the required rate of return on debtholders using the suggested methods. In this case, one can estimate the WACC of a company.

In practice, the market price of liability (debt) equals a sum of payments of the liability discounted at the market interest rate, see, e.g., \citeA{Brealey20}. In this paper, we introduce a simple method that evaluates the market values of liabilities. The method covers not only individual liabilities in the balance sheet but also whole liabilities in the balance sheet.

Our purpose is to estimate the required rate of return on equity. However, the suggested methods can be used to estimate other significant parameters of the private company valuation model. In particular, we estimate price--to--book ratio vector by the ML method with regime--switching and the Bayesian method, and state (unobserved and latent) variable process of price--to--book ratio by the Kalman filtering method. For the Kalman filtering method, we develop the EM algorithm. If we know the book values of the next periods, then one may use forecasting inferences of the state variable to value a company in the next periods. 

Future research works should concentrate on extending the private company valuation model with state variable by state--space model with regime--switching, see \citeA{Kim94}.

\bibliographystyle{apacite}
\bibliography{References}

\begin{thebibliography}{}

\bibitem [\protect \citeauthoryear {%
Agosto%
, Mainini%
\BCBL {}\ \BBA {} Moretto%
}{%
Agosto%
\ \protect \BOthers {.}}{%
{\protect \APACyear {2019}}%
}]{%
Agosto19}
\APACinsertmetastar {%
Agosto19}%
\begin{APACrefauthors}%
Agosto, A.%
, Mainini, A.%
\BCBL {}\ \BBA {} Moretto, E.%
\end{APACrefauthors}%
\unskip\
\newblock
\APACrefYearMonthDay{2019}{}{}.
\newblock
{\BBOQ}\APACrefatitle {Stochastic dividend discount model: covariance of random
  stock prices} {Stochastic dividend discount model: covariance of random stock
  prices}.{\BBCQ}
\newblock
\APACjournalVolNumPages{Journal of Economics and Finance}{43}{3}{552--568}.
\PrintBackRefs{\CurrentBib}

\bibitem [\protect \citeauthoryear {%
Ba{\'n}bura%
, Giannone%
\BCBL {}\ \BBA {} Reichlin%
}{%
Ba{\'n}bura%
\ \protect \BOthers {.}}{%
{\protect \APACyear {2010}}%
}]{%
Banbura10}
\APACinsertmetastar {%
Banbura10}%
\begin{APACrefauthors}%
Ba{\'n}bura, M.%
, Giannone, D.%
\BCBL {}\ \BBA {} Reichlin, L.%
\end{APACrefauthors}%
\unskip\
\newblock
\APACrefYearMonthDay{2010}{}{}.
\newblock
{\BBOQ}\APACrefatitle {Large Bayesian Vector Autoregressions} {Large bayesian
  vector autoregressions}.{\BBCQ}
\newblock
\APACjournalVolNumPages{Journal of Applied Econometrics}{25}{1}{71--92}.
\PrintBackRefs{\CurrentBib}

\bibitem [\protect \citeauthoryear {%
Battulga%
}{%
Battulga%
}{%
{\protect \APACyear {2023}}%
}]{%
Battulga23c}
\APACinsertmetastar {%
Battulga23c}%
\begin{APACrefauthors}%
Battulga, G.%
\end{APACrefauthors}%
\unskip\
\newblock
\APACrefYearMonthDay{2023}{}{}.
\newblock
{\BBOQ}\APACrefatitle {The Log Private Company Valuation Model} {The log
  private company valuation model}.{\BBCQ}
\newblock
\APACjournalVolNumPages{to be appear in Numerical Algebra, Control \&
  Optimization}{}{}{}.
\newblock
\APAChowpublished {Available at: \url{https://arxiv.org/abs/2206.09666}}.
\PrintBackRefs{\CurrentBib}

\bibitem [\protect \citeauthoryear {%
Battulga%
, Jacob%
, Altangerel%
\BCBL {}\ \BBA {} Horsch%
}{%
Battulga%
\ \protect \BOthers {.}}{%
{\protect \APACyear {2022}}%
}]{%
Battulga22a}
\APACinsertmetastar {%
Battulga22a}%
\begin{APACrefauthors}%
Battulga, G.%
, Jacob, K.%
, Altangerel, L.%
\BCBL {}\ \BBA {} Horsch, A.%
\end{APACrefauthors}%
\unskip\
\newblock
\APACrefYearMonthDay{2022}{}{}.
\newblock
{\BBOQ}\APACrefatitle {Dividends and Compound Poisson--Process: A new
  Stochastic Stock Price Model} {Dividends and compound poisson--process: A new
  stochastic stock price model}.{\BBCQ}
\newblock
\APACjournalVolNumPages{International Journal of Theoretical and Applied
  Finance}{25}{3}{2250014}.
\PrintBackRefs{\CurrentBib}

\bibitem [\protect \citeauthoryear {%
Brealey%
, Myers%
\BCBL {}\ \BBA {} Marcus%
}{%
Brealey%
\ \protect \BOthers {.}}{%
{\protect \APACyear {2020}}%
}]{%
Brealey20}
\APACinsertmetastar {%
Brealey20}%
\begin{APACrefauthors}%
Brealey, R.%
, Myers, S\BPBI C.%
\BCBL {}\ \BBA {} Marcus, A\BPBI J.%
\end{APACrefauthors}%
\unskip\
\newblock
\APACrefYear{2020}.
\newblock
\APACrefbtitle {Fundamentals of corporate finance} {Fundamentals of corporate
  finance}\ (\PrintOrdinal{10}\ \BEd).
\newblock
\APACaddressPublisher{}{McGraw Hill}.
\PrintBackRefs{\CurrentBib}

\bibitem [\protect \citeauthoryear {%
Chan%
}{%
Chan%
}{%
{\protect \APACyear {2020}}%
}]{%
Chan20}
\APACinsertmetastar {%
Chan20}%
\begin{APACrefauthors}%
Chan, J\BPBI C.%
\end{APACrefauthors}%
\unskip\
\newblock
\APACrefYearMonthDay{2020}{}{}.
\newblock
{\BBOQ}\APACrefatitle {Large Bayesian vector autoregressions} {Large bayesian
  vector autoregressions}.{\BBCQ}
\newblock
\BIn{} P.~Fuleky\ (\BED), \APACrefbtitle {Macroeconomic Forecasting in the Era
  of Big Data Theory and Practice} {Macroeconomic forecasting in the era of big
  data theory and practice}\ (\BVOL~52, \BPGS\ 95--125).
\newblock
\APACaddressPublisher{}{Springer}.
\PrintBackRefs{\CurrentBib}

\bibitem [\protect \citeauthoryear {%
D'Amico%
\ \BBA {} De~Blasis%
}{%
D'Amico%
\ \BBA {} De~Blasis%
}{%
{\protect \APACyear {2020}}%
{\protect \APACexlab {{\protect \BCnt {1}}}}}]{%
dAmico20b}
\APACinsertmetastar {%
dAmico20b}%
\begin{APACrefauthors}%
D'Amico, G.%
\BCBT {}\ \BBA {} De~Blasis, R.%
\end{APACrefauthors}%
\unskip\
\newblock
\APACrefYearMonthDay{2020{\protect \BCnt {1}}}{}{}.
\newblock
{\BBOQ}\APACrefatitle {A multivariate Markov chain stock model} {A multivariate
  markov chain stock model}.{\BBCQ}
\newblock
\APACjournalVolNumPages{Scandinavian Actuarial Journal}{2020}{4}{272--291}.
\PrintBackRefs{\CurrentBib}

\bibitem [\protect \citeauthoryear {%
D'Amico%
\ \BBA {} De~Blasis%
}{%
D'Amico%
\ \BBA {} De~Blasis%
}{%
{\protect \APACyear {2020}}%
{\protect \APACexlab {{\protect \BCnt {2}}}}}]{%
dAmico20a}
\APACinsertmetastar {%
dAmico20a}%
\begin{APACrefauthors}%
D'Amico, G.%
\BCBT {}\ \BBA {} De~Blasis, R.%
\end{APACrefauthors}%
\unskip\
\newblock
\APACrefYearMonthDay{2020{\protect \BCnt {2}}}{}{}.
\newblock
{\BBOQ}\APACrefatitle {A review of the Dividend Discount Model: from
  deterministic to stochastic models} {A review of the dividend discount model:
  from deterministic to stochastic models}.{\BBCQ}
\newblock
\APACjournalVolNumPages{Statistical Topics and Stochastic Models for Dependent
  Data with Applications}{}{}{47--67}.
\PrintBackRefs{\CurrentBib}

\bibitem [\protect \citeauthoryear {%
Doan%
, Litterman%
\BCBL {}\ \BBA {} Sims%
}{%
Doan%
\ \protect \BOthers {.}}{%
{\protect \APACyear {1984}}%
}]{%
Doan84}
\APACinsertmetastar {%
Doan84}%
\begin{APACrefauthors}%
Doan, T.%
, Litterman, R.%
\BCBL {}\ \BBA {} Sims, C.%
\end{APACrefauthors}%
\unskip\
\newblock
\APACrefYearMonthDay{1984}{}{}.
\newblock
{\BBOQ}\APACrefatitle {Forecasting and conditional projection using realistic
  prior distributions} {Forecasting and conditional projection using realistic
  prior distributions}.{\BBCQ}
\newblock
\APACjournalVolNumPages{Econometric reviews}{3}{1}{1--100}.
\PrintBackRefs{\CurrentBib}

\bibitem [\protect \citeauthoryear {%
Fama%
\ \BBA {} French%
}{%
Fama%
\ \BBA {} French%
}{%
{\protect \APACyear {1993}}%
}]{%
Fama93}
\APACinsertmetastar {%
Fama93}%
\begin{APACrefauthors}%
Fama, E\BPBI F.%
\BCBT {}\ \BBA {} French, K\BPBI R.%
\end{APACrefauthors}%
\unskip\
\newblock
\APACrefYearMonthDay{1993}{}{}.
\newblock
{\BBOQ}\APACrefatitle {Common risk factors in the returns on stocks and bonds}
  {Common risk factors in the returns on stocks and bonds}.{\BBCQ}
\newblock
\APACjournalVolNumPages{Journal of Financial Economics}{33}{1}{3--56}.
\PrintBackRefs{\CurrentBib}

\bibitem [\protect \citeauthoryear {%
Hamilton%
}{%
Hamilton%
}{%
{\protect \APACyear {1989}}%
}]{%
Hamilton89}
\APACinsertmetastar {%
Hamilton89}%
\begin{APACrefauthors}%
Hamilton, J\BPBI D.%
\end{APACrefauthors}%
\unskip\
\newblock
\APACrefYearMonthDay{1989}{}{}.
\newblock
{\BBOQ}\APACrefatitle {A new approach to the economic analysis of nonstationary
  time series and the business cycle} {A new approach to the economic analysis
  of nonstationary time series and the business cycle}.{\BBCQ}
\newblock
\APACjournalVolNumPages{Econometrica: Journal of the Econometric
  Society}{}{}{357--384}.
\PrintBackRefs{\CurrentBib}

\bibitem [\protect \citeauthoryear {%
Hamilton%
}{%
Hamilton%
}{%
{\protect \APACyear {1990}}%
}]{%
Hamilton90}
\APACinsertmetastar {%
Hamilton90}%
\begin{APACrefauthors}%
Hamilton, J\BPBI D.%
\end{APACrefauthors}%
\unskip\
\newblock
\APACrefYearMonthDay{1990}{}{}.
\newblock
{\BBOQ}\APACrefatitle {Analysis of time series subject to changes in regime}
  {Analysis of time series subject to changes in regime}.{\BBCQ}
\newblock
\APACjournalVolNumPages{Journal of Econometrics}{45}{1-2}{39--70}.
\PrintBackRefs{\CurrentBib}

\bibitem [\protect \citeauthoryear {%
Hamilton%
}{%
Hamilton%
}{%
{\protect \APACyear {1994}}%
}]{%
Hamilton94}
\APACinsertmetastar {%
Hamilton94}%
\begin{APACrefauthors}%
Hamilton, J\BPBI D.%
\end{APACrefauthors}%
\unskip\
\newblock
\APACrefYear{1994}.
\newblock
\APACrefbtitle {Time series econometrics} {Time series econometrics}.
\newblock
\APACaddressPublisher{}{Princeton University Press, Princeton}.
\PrintBackRefs{\CurrentBib}

\bibitem [\protect \citeauthoryear {%
Johnston%
\ \BBA {} DiNardo%
}{%
Johnston%
\ \BBA {} DiNardo%
}{%
{\protect \APACyear {1997}}%
}]{%
Johnston97}
\APACinsertmetastar {%
Johnston97}%
\begin{APACrefauthors}%
Johnston, J.%
\BCBT {}\ \BBA {} DiNardo, J.%
\end{APACrefauthors}%
\unskip\
\newblock
\APACrefYear{1997}.
\newblock
\APACrefbtitle {Econometric methods} {Econometric methods}\ (\PrintOrdinal{4}\
  \BEd).
\newblock
\APACaddressPublisher{}{New York}.
\PrintBackRefs{\CurrentBib}

\bibitem [\protect \citeauthoryear {%
Kalman%
}{%
Kalman%
}{%
{\protect \APACyear {1960}}%
}]{%
Kalman60}
\APACinsertmetastar {%
Kalman60}%
\begin{APACrefauthors}%
Kalman, R\BPBI E.%
\end{APACrefauthors}%
\unskip\
\newblock
\APACrefYearMonthDay{1960}{}{}.
\newblock
{\BBOQ}\APACrefatitle {A new approach to linear filtering and prediction
  problems} {A new approach to linear filtering and prediction
  problems}.{\BBCQ}
\newblock
\APACjournalVolNumPages{Journal of Basic Engineering}{82}{1}{35--44}.
\PrintBackRefs{\CurrentBib}

\bibitem [\protect \citeauthoryear {%
Kim%
}{%
Kim%
}{%
{\protect \APACyear {1994}}%
}]{%
Kim94}
\APACinsertmetastar {%
Kim94}%
\begin{APACrefauthors}%
Kim, C\BHBI J.%
\end{APACrefauthors}%
\unskip\
\newblock
\APACrefYearMonthDay{1994}{}{}.
\newblock
{\BBOQ}\APACrefatitle {Dynamic linear models with Markov--switching} {Dynamic
  linear models with markov--switching}.{\BBCQ}
\newblock
\APACjournalVolNumPages{Journal of Econometrics}{60}{1--2}{1--22}.
\PrintBackRefs{\CurrentBib}

\bibitem [\protect \citeauthoryear {%
Leibowitz%
\ \BBA {} Kogelman%
}{%
Leibowitz%
\ \BBA {} Kogelman%
}{%
{\protect \APACyear {1990}}%
}]{%
Leibowitz90}
\APACinsertmetastar {%
Leibowitz90}%
\begin{APACrefauthors}%
Leibowitz, M\BPBI L.%
\BCBT {}\ \BBA {} Kogelman, S.%
\end{APACrefauthors}%
\unskip\
\newblock
\APACrefYearMonthDay{1990}{}{}.
\newblock
{\BBOQ}\APACrefatitle {Inside the P/E Ratio: The Franchise Factor} {Inside the
  p/e ratio: The franchise factor}.{\BBCQ}
\newblock
\APACjournalVolNumPages{Financial Analysts Journal}{46}{6}{17--35}.
\PrintBackRefs{\CurrentBib}

\bibitem [\protect \citeauthoryear {%
L{\"u}tkepohl%
}{%
L{\"u}tkepohl%
}{%
{\protect \APACyear {2005}}%
}]{%
Lutkepohl05}
\APACinsertmetastar {%
Lutkepohl05}%
\begin{APACrefauthors}%
L{\"u}tkepohl, H.%
\end{APACrefauthors}%
\unskip\
\newblock
\APACrefYear{2005}.
\newblock
\APACrefbtitle {New introduction to multiple time series analysis} {New
  introduction to multiple time series analysis}\ (\PrintOrdinal{2}\ \BEd).
\newblock
\APACaddressPublisher{}{Springer Berlin Heidelberg}.
\PrintBackRefs{\CurrentBib}

\bibitem [\protect \citeauthoryear {%
Markowitz%
}{%
Markowitz%
}{%
{\protect \APACyear {1952}}%
}]{%
Markowitz52}
\APACinsertmetastar {%
Markowitz52}%
\begin{APACrefauthors}%
Markowitz, H.%
\end{APACrefauthors}%
\unskip\
\newblock
\APACrefYearMonthDay{1952}{}{}.
\newblock
{\BBOQ}\APACrefatitle {Portfolio Selection} {Portfolio selection}.{\BBCQ}
\newblock
\APACjournalVolNumPages{The Journal of Finance}{7}{1}{77--91}.
\PrintBackRefs{\CurrentBib}

\bibitem [\protect \citeauthoryear {%
McNeil%
, Frey%
\BCBL {}\ \BBA {} Embrechts%
}{%
McNeil%
\ \protect \BOthers {.}}{%
{\protect \APACyear {2005}}%
}]{%
McNeil05}
\APACinsertmetastar {%
McNeil05}%
\begin{APACrefauthors}%
McNeil, A\BPBI J.%
, Frey, R.%
\BCBL {}\ \BBA {} Embrechts, P.%
\end{APACrefauthors}%
\unskip\
\newblock
\APACrefYear{2005}.
\newblock
\APACrefbtitle {Quantitative risk management: concepts, techniques and tools}
  {Quantitative risk management: concepts, techniques and tools}.
\newblock
\APACaddressPublisher{}{Princeton University Press}.
\PrintBackRefs{\CurrentBib}

\bibitem [\protect \citeauthoryear {%
Nagorniak%
}{%
Nagorniak%
}{%
{\protect \APACyear {1985}}%
}]{%
Nagorniak85}
\APACinsertmetastar {%
Nagorniak85}%
\begin{APACrefauthors}%
Nagorniak, J.%
\end{APACrefauthors}%
\unskip\
\newblock
\APACrefYearMonthDay{1985}{}{}.
\newblock
{\BBOQ}\APACrefatitle {Thoughts on using dividend discount models} {Thoughts on
  using dividend discount models}.{\BBCQ}
\newblock
\APACjournalVolNumPages{Financial Analysts Journal}{41}{6}{13--15}.
\PrintBackRefs{\CurrentBib}

\bibitem [\protect \citeauthoryear {%
Ross%
}{%
Ross%
}{%
{\protect \APACyear {1976}}%
}]{%
Ross76}
\APACinsertmetastar {%
Ross76}%
\begin{APACrefauthors}%
Ross, S\BPBI A.%
\end{APACrefauthors}%
\unskip\
\newblock
\APACrefYearMonthDay{1976}{}{}.
\newblock
{\BBOQ}\APACrefatitle {The arbitrage theory of capital asset pricing} {The
  arbitrage theory of capital asset pricing}.{\BBCQ}
\newblock
\APACjournalVolNumPages{Journal of Economic Theory}{13}{3}{341--360}.
\PrintBackRefs{\CurrentBib}

\bibitem [\protect \citeauthoryear {%
Schneider%
}{%
Schneider%
}{%
{\protect \APACyear {1992}}%
}]{%
Schneider92}
\APACinsertmetastar {%
Schneider92}%
\begin{APACrefauthors}%
Schneider, W.%
\end{APACrefauthors}%
\unskip\
\newblock
\APACrefYearMonthDay{1992}{}{}.
\newblock
{\BBOQ}\APACrefatitle {Systems of seemingly unrelated regression equations with
  time varying coefficients—An interplay of Kalman filtering, scoring, EM-and
  MINQUE-method} {Systems of seemingly unrelated regression equations with time
  varying coefficients—an interplay of kalman filtering, scoring, em-and
  minque-method}.{\BBCQ}
\newblock
\APACjournalVolNumPages{Computers \& Mathematics with
  Applications}{24}{8-9}{1--16}.
\PrintBackRefs{\CurrentBib}

\bibitem [\protect \citeauthoryear {%
Williams%
}{%
Williams%
}{%
{\protect \APACyear {1938}}%
}]{%
Williams38}
\APACinsertmetastar {%
Williams38}%
\begin{APACrefauthors}%
Williams, J\BPBI B.%
\end{APACrefauthors}%
\unskip\
\newblock
\APACrefYear{1938}.
\newblock
\APACrefbtitle {The theory of investment value} {The theory of investment
  value}.
\newblock
\APACaddressPublisher{}{Harvard University Press}.
\PrintBackRefs{\CurrentBib}

\bibitem [\protect \citeauthoryear {%
Zucchini%
, MacDonald%
\BCBL {}\ \BBA {} Langrock%
}{%
Zucchini%
\ \protect \BOthers {.}}{%
{\protect \APACyear {2016}}%
}]{%
Zucchini16}
\APACinsertmetastar {%
Zucchini16}%
\begin{APACrefauthors}%
Zucchini, W.%
, MacDonald, I\BPBI L.%
\BCBL {}\ \BBA {} Langrock, R.%
\end{APACrefauthors}%
\unskip\
\newblock
\APACrefYear{2016}.
\newblock
\APACrefbtitle {Hidden Markov models for time series: an introduction using R}
  {Hidden markov models for time series: an introduction using r}\
  (\PrintOrdinal{2}\ \BEd).
\newblock
\APACaddressPublisher{}{CRC press}.
\PrintBackRefs{\CurrentBib}

\end{thebibliography}

\end{document}